\documentclass[aps,prl,twocolumn,floats,balancelastpage,showpacs,showkeys,preprintnumbers,floatfix,nofootinbib,superscriptaddress]{revtex4-1}
\usepackage{graphicx,amsmath,amssymb,amsfonts, soul, amssymb,color,float,wasysym,wrapfig,xspace,changepage,todonotes}
\usepackage[colorlinks]{hyperref}

\begin{document}
\title{Search for dark matter annihilation signals in the H.E.S.S. Inner Galaxy Survey}

\author{H.E.S.S. Collaboration}
%\email[Corresponding authors: D. Malyshev, A. Montanari, E. Moulin\\ Correspondence should be sent to ]{contact.hess@hess-experiment.eu}
\noaffiliation

\author{H.~Abdalla}
\affiliation{University of Namibia, Department of Physics, Private Bag 13301, Windhoek 10005, Namibia}

\author{F.~Aharonian}
\affiliation{Dublin Institute for Advanced Studies, 31 Fitzwilliam Place, Dublin 2, Ireland}
\affiliation{Max-Planck-Institut f\"ur Kernphysik, P.O. Box 103980, D 69029 Heidelberg, Germany}
\affiliation{High Energy Astrophysics Laboratory, RAU,  123 Hovsep Emin St  Yerevan 0051, Armenia}

\author{F.~Ait~Benkhali}
\affiliation{Max-Planck-Institut f\"ur Kernphysik, P.O. Box 103980, D 69029 Heidelberg, Germany}

\author{E.O.~Ang\"uner}
\affiliation{Aix Marseille Universit\'e, CNRS/IN2P3, CPPM, Marseille, France}

\author{C.~Armand}
\affiliation{Université Savoie Mont Blanc, CNRS, Laboratoire d'Annecy de Physique des Particules - IN2P3, 74000 Annecy, France}

%\author[0000-0002-2153-1818]{H.~Ashkar}
\author{H.~Ashkar}
\affiliation{IRFU, CEA, Universit\'e Paris-Saclay, F-91191 Gif-sur-Yvette, France}

%\author[0000-0002-9326-6400]{M.~Backes}
\author{M.~Backes}
\affiliation{University of Namibia, Department of Physics, Private Bag 13301, Windhoek 10005, Namibia}
\affiliation{Centre for Space Research, North-West University, Potchefstroom 2520, South Africa}

\author{V.~Baghmanyan}
\affiliation{Instytut Fizyki J\c{a}drowej PAN, ul. Radzikowskiego 152, 31-342 Krak{\'o}w, Poland}

%\author[0000-0002-5085-8828]{V.~Barbosa~Martins}
\author{V.~Barbosa~Martins}
\affiliation{DESY, D-15738 Zeuthen, Germany}

% ...

%\collaboration{1}{H.E.S.S. Collaboration}

\begin{abstract}
The central region of the Milky Way is one of the foremost locations to look for dark matter (DM) signatures. 
We report the first results on a search for DM particle annihilation signals using new observations  from an unprecedented gamma-ray survey of the Galactic Center (GC) region, {\it i.e.}, the Inner Galaxy Survey, at very high energies ($\gtrsim$ 100 GeV) performed with the H.E.S.S. array of five ground-based Cherenkov telescopes. No significant gamma-ray excess is found in the search region of the 2014-2020 dataset and a profile likelihood ratio analysis is carried out to set exclusion limits on the annihilation cross section $\langle \sigma v\rangle$.
Assuming Einasto and Navarro-Frenk-White (NFW) DM density profiles at the GC, these constraints are the strongest obtained so far in the TeV DM mass range. For the Einasto profile, the constraints reach $\langle \sigma v\rangle$ values of $\rm 3.7\times10^{-26} cm^3s^{-1}$ for 1.5 TeV DM mass 
in the $W^+W^-$ annihilation channel, and $\rm 1.2 \times 10^{-26} cm^3s^{-1}$ for 0.7 TeV DM mass in the $\tau^+\tau^-$ annihilation channel.
With the H.E.S.S. Inner Galaxy Survey,  ground-based $\gamma$-ray observations thus probe $\langle \sigma v\rangle$ values expected from thermal-relic annihilating TeV DM particles.
\end{abstract}


\maketitle 

\section{Introduction}
\label{sec:introduction} 
The total matter content of the Universe is made of about 85\% dark and non-baryonic matter as suggested by growing evidence from astrophysics and cosmology~\cite{Planck:2015fie,Planck:2018vyg}. However, the nature of the dark matter (DM) remains a fundamental question of modern physics. A compelling class of stable DM candidates are weakly interacting massive 
particles (WIMPs)~\cite{Bertone:2010zza,Feng:2010gw,Roszkowski:2017nbc}.  Such particles with mass and coupling strength at the electroweak scale naturally emerges in several extensions of the standard model of particle physics. WIMPs are thermally produced in the early Universe 
and their relic density can represent all the DM in the Universe~\cite{Jungman:1995df}, as accurately measured from cosmological observations. The quest for WIMPs motivates numerous  
experimental efforts to probe their non-gravitational properties such as
their production at particle colliders~\cite{Kahlhoefer:2017dnp}, their scattering off nuclei on Earth~\cite{Schumann:2019eaa}, and their decay and annihilation~\cite{Strigari:2018utn}.

WIMPs would self-annihilate today in dense astrophysical environments, producing gamma rays in the final state from hadronization, radiation and decay of the standard model particles produced in the annihilation process. Such gamma rays could be eventually detected by ground-based arrays of Imaging Atmospheric Cherenkov Telescopes (IACTs) such as the High Energy Stereoscopic System (H.E.S.S.) provided that the WIMP mass is high enough.
The self-annihilation of Majorana WIMPs of mass $m_{\rm DM}$ would produce an energy-differential flux of gamma rays in a solid angle $\Delta\Omega$ expressed as:
\begin{widetext}
\begin{equation}
\label{eq:dmflux}
\frac{\rm d \Phi_\gamma}{\rm d E_\gamma} (E_\gamma,\Delta\Omega)=
\frac{\langle \sigma v \rangle}{8\pi m_{\rm DM}^2}\sum_f  BR_f \frac{d N^f_\gamma}{d E_\gamma}(E_\gamma) \, J(\Delta\Omega) 
\quad {\rm with} \quad  J(\Delta\Omega) =  \int_{\Delta\Omega} \int_{\rm los} \rho^2(s(r,\theta)) ds\, d\Omega \, .
\end{equation}
\end{widetext}
$\langle \sigma v \rangle$ is the velocity-weighted annihilation cross section averaged over the velocity distribution and $dN^f_\gamma/dE_\gamma$ is the differential yield of gamma rays per annihilation in the channel $f$ with its branching ratio $BR_{f}$. The term $J(\Delta\Omega)$, hereafter referred to as the J-factor, corresponds to the integral of the square of the DM density $\rho$ over the line of sight (los) $s$ and solid angle $\Delta\Omega$. The DM density $\rho$ is assumed spherically symmetric and therefore depends only on the radial coordinate $r$ from the center of the DM halo.  
It can be expressed as $r = \big(s^2 +r_{\odot}^2-2\,r_{\odot}\,s\, \cos\theta \big)^{1/2}$, with $r_\odot$ is the distance of the observer to the GC taken to be
$r_\odot$ = 8.5 kpc~\cite{Ghez:2008ms}, and $\theta$ is the angle between the direction of observation and the Galactic Center.
The centre of the Milky Way is predicted as the brightest source of DM annihilations with a DM distribution assumed to follow cuspy profiles conveniently described by the Einasto~\cite{Springel:2008by} or Navarro-Frenk-White~\cite{Navarro:1996gj} parametrizations. Commonly-used sets of parameters for the above-mentioned DM  profiles~\cite{Abdallah:2016ygi,Abdallah:2018qtu} considered here are given in 
Tab.~II of Ref.~\cite{supplement}. 
The DM profiles are normalized to the local DM density $\rho_\odot$ such that 
$\rho(r_\odot) = \rho_\odot$ = 0.39 GeVcm$^{-3}$~\cite{Catena:2009mf}.
Improved determinations of the local DM density are being carried out (see, for instance, Ref.~\cite{Read:2014qva})\footnote{Estimates of the local DM density show an uncertainty of about a factor of 2~\cite{Zyla:2020zbs}.}. A change of $\rho_\odot$ can be propagated to the results by rescaling the DM signal by $(\rho_\odot$/0.39 GeVcm$^{-3})^2$.
Other parameterizations such as the Burkert~\cite{Burkert:1995yz} or Moore~\cite{Diemand:2004wh} profile can be used. However, cored profiles such as the Burkert one are not studied here, since they need dedicated observations and analysis procedures~\cite{HESS:2015cda}.
The strongest constraints so far obtained on WIMPs in the TeV mass range come from 254 hours of H.E.S.S. observations of the Galactic Center region~\cite{Abdallah:2016ygi}. In the present work, about 5 times more exposure in total is available with respect to the previous H.E.S.S. observations~\cite{supplement}.

In this Letter, we report on a new search for DM annihilation in the central region of the Milky Way halo using an unprecedented dataset from very-high-energy (VHE, E $\gtrsim$ 100 GeV) observations taken with the five-telescope H.E.S.S. array of the Galactic Centre region.

\section{Observations and data analysis}
\label{sec:analysis}
The H.E.S.S. collaboration is carrying out an extensive observation programme to survey the central region of the Milky Way. Such a region can be observed with the H.E.S.S. 
observatory under very good conditions due to
its location near the tropic of Capricorn.
The survey aims at covering the inner several hundred parsecs of the Galactic Center region, in order to achieve the best possible sensitivity for DM annihilation signals and Galactic Center outflows. 
Such a survey, hereafter referred to as the Inner Galaxy Survey (IGS), is the first-ever conducted deep VHE gamma-ray survey of the Galactic Center region.
In order to cover so far unexplored regions in VHE gamma rays, the current implementation of the IGS is based on the definition of a grid of telescope pointing positions up to Galactic latitudes $b$ = +3.2$^\circ$, 
as shown in the top-left panel of 
Fig.~1 of Ref.~\cite{supplement}.
The present dataset makes use of 28-minute data taking runs between 2014 and 2020, amounting to a total of 546 hours (live time) of high-quality data following the standard data quality selection procedure~\cite{Aharonian:2006pe}. Observations are taken at observational zenith angles lower than 40$^\circ$ to minimize systematic uncertainties in the event reconstruction. The observational campaign results in an averaged observational zenith angle of 18$^\circ$ for the present dataset. 
An acceptance-corrected exposure time of at least ten hours is reached up to $b \approx$ +6$^\circ$ with the present dataset. 
Gamma-ray-like events are selected and reconstructed with a semi-analytical shower model technique
based on a fit of observed shower images to a semi-analytical shower model~\cite{2009APh32231D}.
An angular resolution of 0.06$^\circ$ (68\% containment radius) and an energy resolution of 10\% above 200 GeV are achieved.
The central region of the Milky Way is a complex environment including numerous regions with VHE gamma-ray emission~\cite{Aharonian:2009zk,Abramowski:2016mir,H.E.S.S.:2018zkf} as well as varying night sky background in the field of view~\cite{Abdallah:2018qtu}. A study of the systematic uncertainties in the background determination is presented in Ref.~\cite{supplement}. 

The DM annihilation signal is searched in regions of interest (ROI) defined as rings centered on the nominal GC position. In order to avoid gamma-ray contamination from known astrophysical sources in the whole field of view, a conservative set of exclusion regions is defined (see
Fig.~1 in Ref.~\cite{supplement})  according to the H.E.S.S. angular resolution and the extension of the emissions in the field of view. See Ref.~\cite{supplement} for more details. The ROIs are therefore considered with inner radii from 0.5$^\circ$ to 2.9$^\circ$, and width of 0.1$^\circ$ each. This set of 25 rings is hereafter referred as to the ON region. 
For each ROI, the residual gamma-ray  background is measured on a run-by-run basis in a region of the field of view taken symmetrically to the ON region from the pointing position, which is hereafter referred to as the OFF region.
The excluded regions are similarly removed from the ON and OFF regions such that they keep the same solid angle and acceptance. 
The OFF regions are always sufficiently far away from the ON regions, such that a significant difference in the expected DM signal between ON and OFF regions is obtained.
More details are provided in 
Fig.~2 of Ref.~\cite{supplement}.
 Any potential unaccounted gamma-ray emission is considered as part
of the measured excess, which makes the analysis conservative as long as no signal is detected.

For each ROI, event distributions are built as a function of energy and are hereafter referred to as the energy count distributions. The systematic uncertainty on the normalisation of the measured energy count distributions is 1\%~\cite{supplement}. 
\begin{figure*}[htp!]
\includegraphics[width=0.45\textwidth]{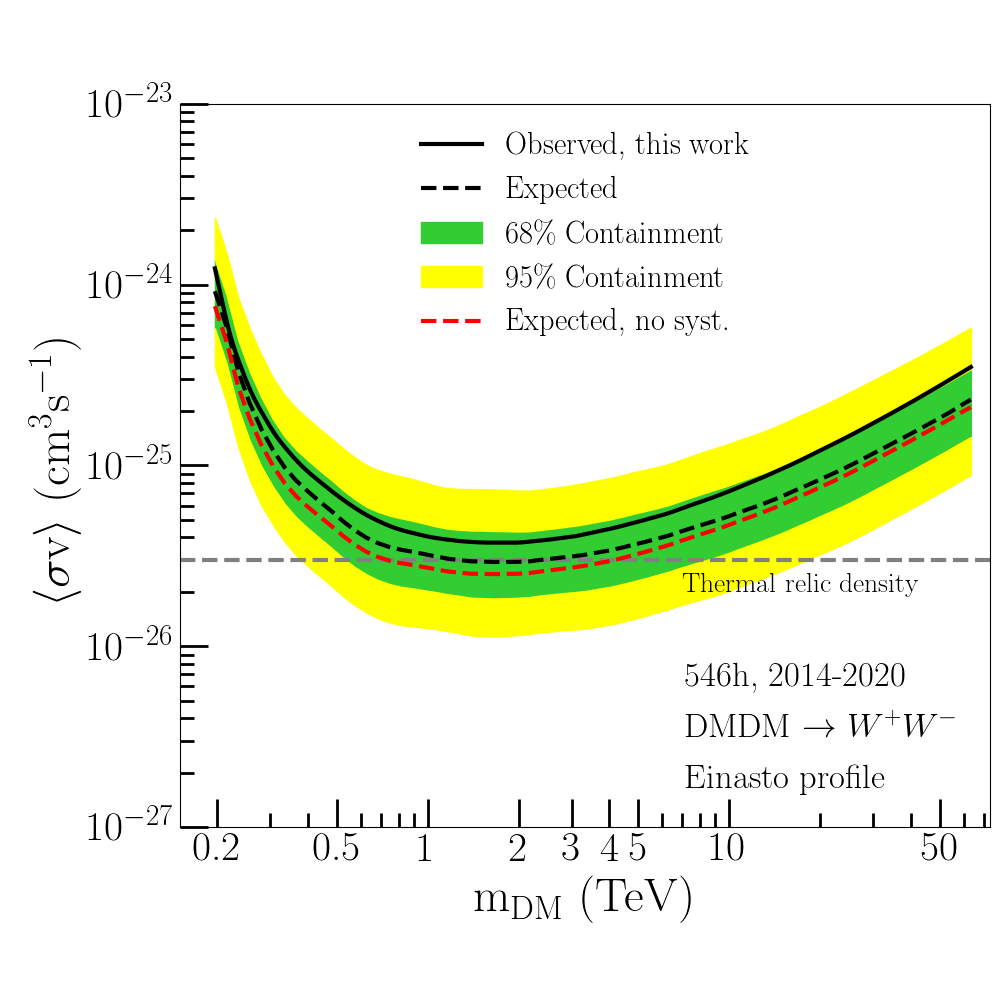}
\includegraphics[width=0.45\textwidth]{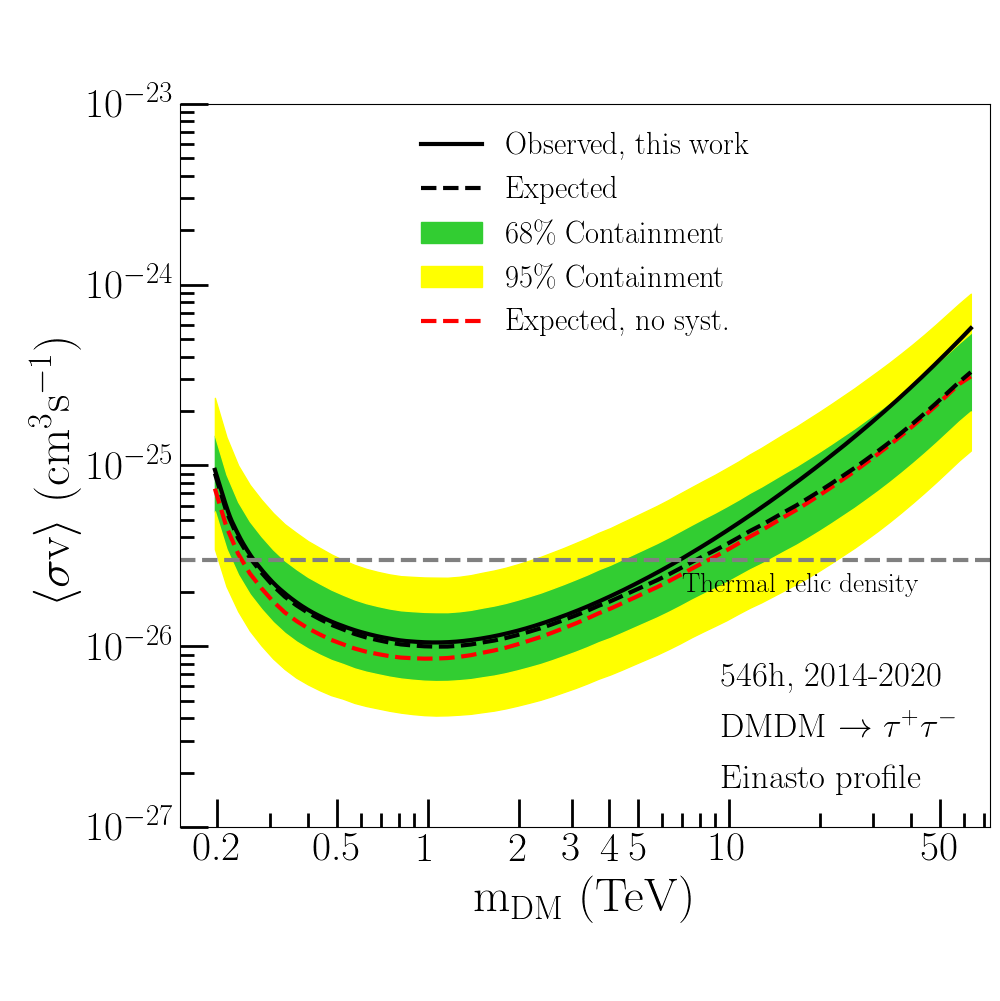}
\caption{Constraints on the velocity-weighted annihilation cross section $\langle \sigma v \rangle$ 
for the  $W^+W^-$ (left panel) and  $\tau^+\tau^-$ (right panel) channels derived from the H.E.S.S. observations taken from 2014 to 2020. 
The constraints are expressed as 95\% C. L. upper limits including the systematic uncertainty on $\langle \sigma v \rangle$ as a function of the DM mass $m_{\rm DM}$. 
 The observed limit is shown as black solid line. 
 The mean expected limit (black dashed line) together with the 68\% (green band) and 95\% (yellow band) 
 C.L. statistical containment bands are shown.
The mean expected upper limit without systematic uncertainty is also shown (red dashed line).
The horizontal grey long-dashed line is set to the value of the natural scale expected for the thermal-relic WIMPs.
The constraints obtained in the $b\bar{b}$, $t\bar{t}$, $ZZ$, $hh$, $\mu^+\mu^-$ and $e^+e^-$ channels are given in 
Fig.~3 of Ref.~\cite{supplement}.}
\label{fig:results}
\end{figure*}

The statistical data analysis is based on a 2-dimensional log-likelihood ratio test statistic which makes use of the expected spectral and spatial DM signal features in 67 logarithmically-spaced energy bins and 25 spatial bins corresponding to the ROI. For a given DM mass, the likelihood function reads:
\begin{widetext}
\begin{equation}
\mathcal{L}_{\rm ij}({\bf N}^{\rm S},{\bf N}^{\rm B}|{\bf N}_{\rm ON},{\bf N}_{\rm OFF}) = \frac{[\beta_{\rm ij}(N_{\rm ij}^{\rm S}+N_{\rm ij}^{\rm B})]^{N_{{\rm ON, ij}}}}{N_{{\rm ON, ij}}!}e^{-\beta_{\rm ij}(N_{\rm ij}^{\rm S}+ N_{\rm ij}^{\rm B})} 
\frac{[\beta_{\rm ij}(N_{\rm ij}^{\rm S'}+N_{\rm ij}^{\rm B})]^{N_{{\rm OFF, ij}}}}{N_{{\rm OFF, ij}}!}e^{-\beta_{\rm ij}(N_{\rm ij}^{\rm S'}+N_{\rm ij}^{\rm B})} e^{-\frac{(1-\beta_{\rm ij})^2}{2\sigma_{\beta_{\rm ij}}^2}} \, .
\label{eq:lik}
\end{equation}
\end{widetext}
$N_{\rm ON,ij}$ and $N_{\rm OFF,ij}$ are the number of measured events in the ON and OFF regions, respectively, in the spectral bin  $i$ and in the spatial bin $j$.
$N^{\rm B}_{\rm ij}$ is the expected number of background events in the $(i,j)$ bin for the ON and OFF regions.
$N^{\rm S}_{\rm ij}$ and $N^{\rm S'}_{\rm ij}$ are the total number of DM events in the $(i,j)$ bin for the ON and OFF regions, respectively. 
It is obtained by folding the expected DM flux given in Eq.(\ref{eq:dmflux}) with the energy-dependent acceptance and energy resolution. The gamma-ray yield $dN^f_\gamma/dE_\gamma$ in the channel $f$ is computed with the Monte Carlo event collision generator
PYTHIAv8.135, including final state radiative corrections~\cite{Cirelli:2010xx}.
The J-factor values of each ROI are reported
in Tab.~III of 
Ref.~\cite{supplement}. $N^{\rm S}_{\rm ij} + N^{\rm B}_{\rm ij}$ is the total number of events in the spatial bin $j$ and spectral bin $i$. The systematic uncertainty can be accounted for in the likelihood function as a Gaussian nuisance parameter where $\beta_{\rm ij}$  acts as a normalisation factor and 
 $\sigma_{\beta_{\rm ij}}$ is the width of the Gaussian function (see, for instance, Refs.~\cite{Silverwood:2014yza,Lefranc:2015pza,Moulin:2019oyc}). $\beta_{\rm ij}$ is found by maximizing the likelihood function such that $\rm d
\mathcal{L}_{\rm ij}/d \beta_{\rm ij} \equiv 0$. A value of 1\% for $\sigma_{\beta_{\rm ij}}$ is used~\cite{supplement}.

In case of no significant excess in the ROIs, constraints on $\langle \sigma v \rangle$ are obtained from the  log-likelihood ratio TS described in Ref.~\cite{2011EPJC711554C} assuming
a positive signal 
$\langle \sigma v \rangle >$~0~\cite{supplement}. We used the high statistics limit in which the TS follows a $\chi^2$ distribution with one degree of freedom. Values of $\langle \sigma v \rangle$ for which TS is higher than 2.71 are excluded at the 95\% confidence level (C.L.).

\section{Results}
\label{sec:results}
We find no significant excess in any of the ON regions with respect to the OFF regions. An analysis crosscheck performed using independent event calibration and reconstruction~\cite{Parsons:2014voa} corroborates the absence of significant excess.
Hence, we derive 95\% C.L. upper limits on $\langle \sigma v \rangle$. We explore the self-annihilation of WIMPs with masses from 200 GeV up to 70 TeV, into the quark ($b\bar{b}$, $t\bar{t}$), gauge bosons ($W^+W^-$, $ZZ$), lepton ($e^+e^-$, $\mu^+\mu^-$, $\tau^+\tau^-$) and Higgs ($hh$) channels, respectively.

\begin{figure*}[ht!]
\includegraphics[width=0.45\textwidth]{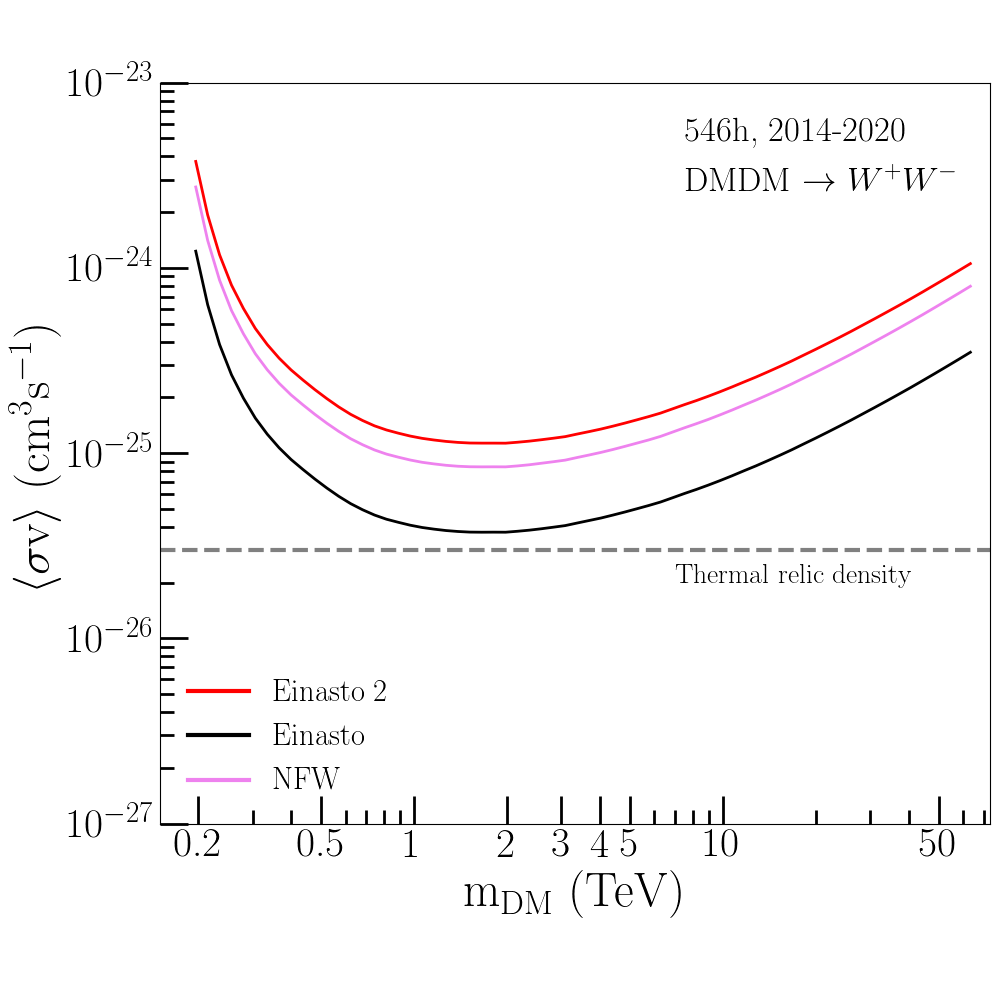}
\includegraphics[width=0.45\textwidth]{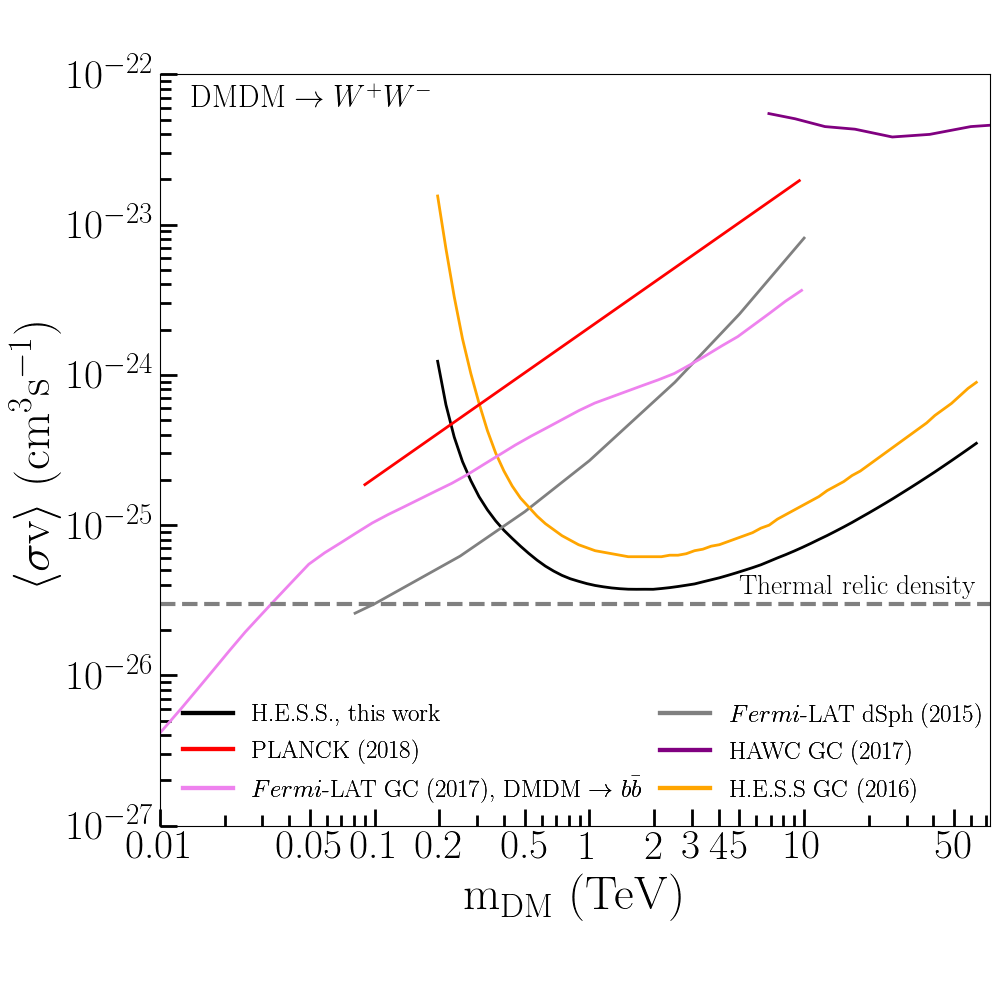}
\caption{{\it Left panel:} Impact of the DM density distribution on the constraints on the velocity-weighted annihilation cross section $\langle \sigma v \rangle$. The constraints expressed in terms of 95\% C. L. upper limits including the systematic uncertainty,
are shown as a function of the DM mass $m _{\rm DM}$ in the $W^+W^-$ channel
for the Einasto profile (black line), another parametrization of the Einasto profile~\cite{Cirelli:2010xx} referred as to Einasto 2 (red line), and the NFW profile (pink line), respectively.
{\it Right panel:} Comparison of present constraints in the $W^+W^-$ channel with the previous published H.E.S.S. limits from 254 hours of observations of the GC~\cite{Abdallah:2016ygi} (orange line), the limits from the observation of the GC with HAWC~\cite{Abeysekara_2018} (purple line), the limits from the observations of 15 dwarf galaxy satellites of the Milky Way by the Fermi satellite~\cite{Ackermann:2015zua} (grey line), the limits from the cosmic microwave background with PLANCK~\cite{Planck:2018vyg} (red line). 
The limits from the observation of the GC with the Fermi satellite in the $b\bar{b}$ channel~\cite{Fermi-LAT:2017opo} are also shown (violet line). The Einasto profile is used for GC observations.}
\label{fig:SummaryPlot}
\end{figure*}
Fig.~\ref{fig:results} shows the 95\% C.L.  observed and expected upper limits for the $W^+W^-$ and $\tau^+\tau^-$ channels, respectively, for the above-mentioned Einasto profile. The observed limits are computed with ON and OFF measured event distributions. The expected limits are obtained from 300 Poisson realizations of the background extracted from the OFF regions. See Supplement Material~\cite{supplement} for more details.
The mean expected upper limit and the 68\% and 95\% containment bands are plotted. 
The 95\% C.L. observed limits reach $\rm 3.7\times10^{-26} cm^3s^{-1}$ for a DM particle mass of 1.5 TeV in the $W^+W^-$ channel, and $\rm 1.2 \times 10^{-26} cm^3s^{-1}$ for 0.7 TeV DM mass in the $\tau^+\tau^-$ annihilation channel. 
The limits in the $\tau^+\tau^-$ annihilation channel cross the $\langle \sigma v \rangle$ values expected for DM particles annihilating with thermal-relic cross section~\cite{Bertone:2004pz}. The limits for the other annihilation channels are shown in Fig.~3 of Ref.~\cite{supplement}. 
At 1.5 TeV DM mass, we obtain 
an improvement factor of 1.6
with respect to the results shown in 
Ref.~\cite{Abdallah:2016ygi}.
The larger statistics of the dataset from longer observational live time and the data taking with the CT1-5 array of H.E.S.S. contribute to the higher sensitivity of the present analysis.

The left panel of Fig.~\ref{fig:SummaryPlot} shows the limits for the NFW profile as well as an alternative set of parameters for the Einasto profile described in Ref.~\cite{Cirelli:2010xx}.
Assuming a kiloparsec-sized cored DM density distribution such as the Burkert profile would weaken the limits by about two orders of magnitude, while a Moore-like profile would improve the limit by a factor of about two.

The right panel of Fig.~\ref{fig:SummaryPlot} summarizes the 
limits obtained from 254 hours of previous H.E.S.S. observation of the Galactic Center~\cite{Abdallah:2016ygi}, from the HAWC observation of the Galactic Center~\cite{Abeysekara_2018}, from the observation of 15 dwarf galaxy satellites of the Milky Way~\cite{Ackermann:2015zua} as well as from the observation of the GC with $Fermi$-LAT~\cite{Fermi-LAT:2017opo}, and the limits from the cosmic microwave background measured by PLANCK~\cite{Planck:2018vyg}. The present H.E.S.S. constraints surpass the {\it Fermi}-LAT limits for particle masses above $\sim$300 GeV.

\section{Summary}
In this Letter we report on the latest results 
on a search for annihilating DM signals from new observations of the inner halo of the Milky Way with the H.E.S.S. five-telescope array. The present dataset amounts to 546 h of total live time spread over 6 years of H.E.S.S. observations. 
The absence of significant excess yields constraints on the velocity-weighted annihilation cross section of Majorana WIMPs. In the $W^+W^-$ channel we obtain 95\% C.L. upper limits of $\rm 3.7\times10^{-26} cm^3s^{-1}$ for DM particles with mass of 1.5 TeV, assuming an Einasto profile. These new limits improve significantly upon the previous constraints and are the most constraining so far in the TeV mass range. The strongest limits are obtained for the $\tau^+\tau^-$ channel, reaching $\rm 1.2 \times 10^{-26} cm^3s^{-1}$, for a DM particle mass of 0.7 TeV. The limits obtained in the $\tau^+\tau^-$ and $e^+e^-$ channels 
challenge natural $\langle \sigma v \rangle$ values expected for the thermal-relic WIMPs in the TeV DM mass range.  
The observations carried out with the IGS program as well as the use of the full five-telescope array contribute to the improved sensitivity of this analysis. 
VHE observations of the central region of the Milky Way with IACTs such as H.E.S.S. are unique for an in-depth study of WIMP models and provide a crucial insight of the TeV WIMP DM paradigm. 
They provide an unprecedented dataset to explore the yet-uncharted parameter space of multi-TeV DM models such as the benchmark candidates Wino and Higgsino (see, for instance, Ref.~\cite{Rinchiuso:2020skh} and references therein) which naturally arise in simple extensions to the Standard Model. The IGS program carried out with H.E.S.S. is an important legacy of H.E.S.S. and paves the way to future Southern-site observations with CTA~\cite{Moulin:2019oyc}.

%%%%%%%%%%%%%%%%%%%%%%%%%%%%%%%%%%%%%%%%%%%%%%%%%%%%%%%%%%%%%%%%%%%%%%%
\section{Acknowledgements}
%%%%%%%%%%%%%%%%%%%%%%%%%%%%%%%%%%%%%%%%%%%%%%%%%%%%%%%%%%%%%%%%%%%%%%%
The support of the Namibian authorities and of the University of Namibia in facilitating the construction and operation of H.E.S.S. is gratefully acknowledged, as is the support by the German Ministry for Education and Research (BMBF), the Max Planck Society, the German Research Foundation (DFG), the Helmholtz Association, the Alexander von Humboldt Foundation, the French Ministry of Higher Education, Research and Innovation, the Centre National de la Recherche Scientifique (CNRS/IN2P3 and CNRS/INSU), the Commissariat \`a l'\'energie atomique et aux \'energies alternatives (CEA), the U.K. Science and Technology Facilities Council (STFC), the Knut and Alice Wallenberg Foundation, the National Science Centre, Poland grant no. 2016/22/M/ST9/00382, the South African Department of Science and Technology and National Research Foundation, the University of Namibia, the National Commission on Research, Science \& Technology of Namibia (NCRST), the Austrian Federal Ministry of Education, Science and Research and the Austrian Science Fund (FWF), the Australian Research Council (ARC), the Japan Society for the Promotion of Science and by the University of Amsterdam.

We appreciate the excellent work of the technical support staff in Berlin, Zeuthen, Heidelberg, Palaiseau, Paris, Saclay, T\"ubingen and in Namibia in the construction and operation of the equipment. This work benefitted from services provided by the H.E.S.S. Virtual Organisation, supported by the national resource providers of the EGI Federation.

\bibliography{bibl}

\clearpage
\appendix
\setcounter{equation}{0}
\setcounter{figure}{0}
\widetext
\begin{center}
{\bf \large \large Supplemental Material: Search for dark matter annihilation signals in the H.E.S.S. Inner Galaxy Survey}
\end{center}
% \documentclass[aps,prl,onecolumn,floats,balancelastpage,showpacs,showkeys,preprintnumbers,floatfix,nofootinbib,superscriptaddress,linenumbers]{revtex4-1}
% \usepackage{graphicx,amsmath,amssymb,amsfonts, soul, amssymb,color,float,wasysym,wrapfig,xspace,changepage,lineno,todonotes}
% \usepackage[colorlinks]{hyperref}

% \begin{document}
\title{Supplemental Material: Search for dark matter annihilation signals with the H.E.S.S. Inner Galaxy Survey}

\maketitle 

\section{Pointing positions of the telescopes for the 
Inner Galaxy Survey}
The H.E.S.S. collaboration is carrying out an extensive observation campaign to survey the central region of the Galactic halo within several degrees from the Galactic Centre. Such a survey is hereafter referred to as the Inner Galaxy Survey (IGS). The IGS is a multiple-year observation program with the five-telescope array of H.E.S.S. The IGS started in 2016, 
covering the Galactic Centre region with significant exposure at Galactic longitudes $|l|<$ 5$^\circ$ and latitudes $b$ from -3$^\circ$ to 6$^\circ$. The data set used in the present study also includes earlier dedicated observations  towards the supermassive black hole Sagittarius A$^*$ taken in 2014 and 2015.
The pointing positions at that time 
were chosen for the needs of the  Galactic plane survey~\cite{H.E.S.S.:2018zkf} and dedicated source observations such as the pulsar PSR J1723-2837.

Given the well-suited location of the H.E.S.S. instrument to observe the GC region, the aim of such a survey is to reach the best sensitivity to diffuse VHE emissions with the lowest energy threshold. To this purpose, observations are taken with the five telescopes under nominal darkness and very good atmospheric conditions with a zenith angle lower than 40$^\circ$. The IGS dataset used in this work is obtained from 1317 observational runs. 
\begin{figure}[!ht]
\includegraphics[width=0.45\textwidth]{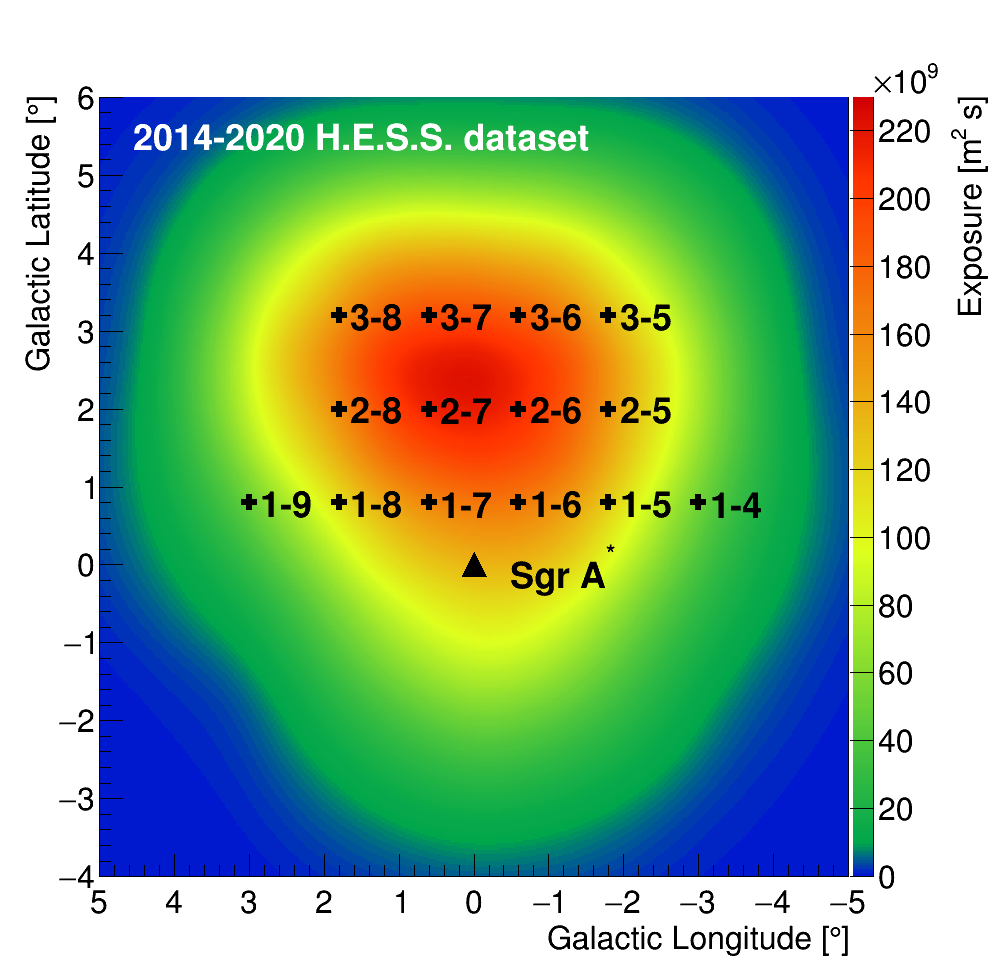}
\includegraphics[width=0.45\textwidth]{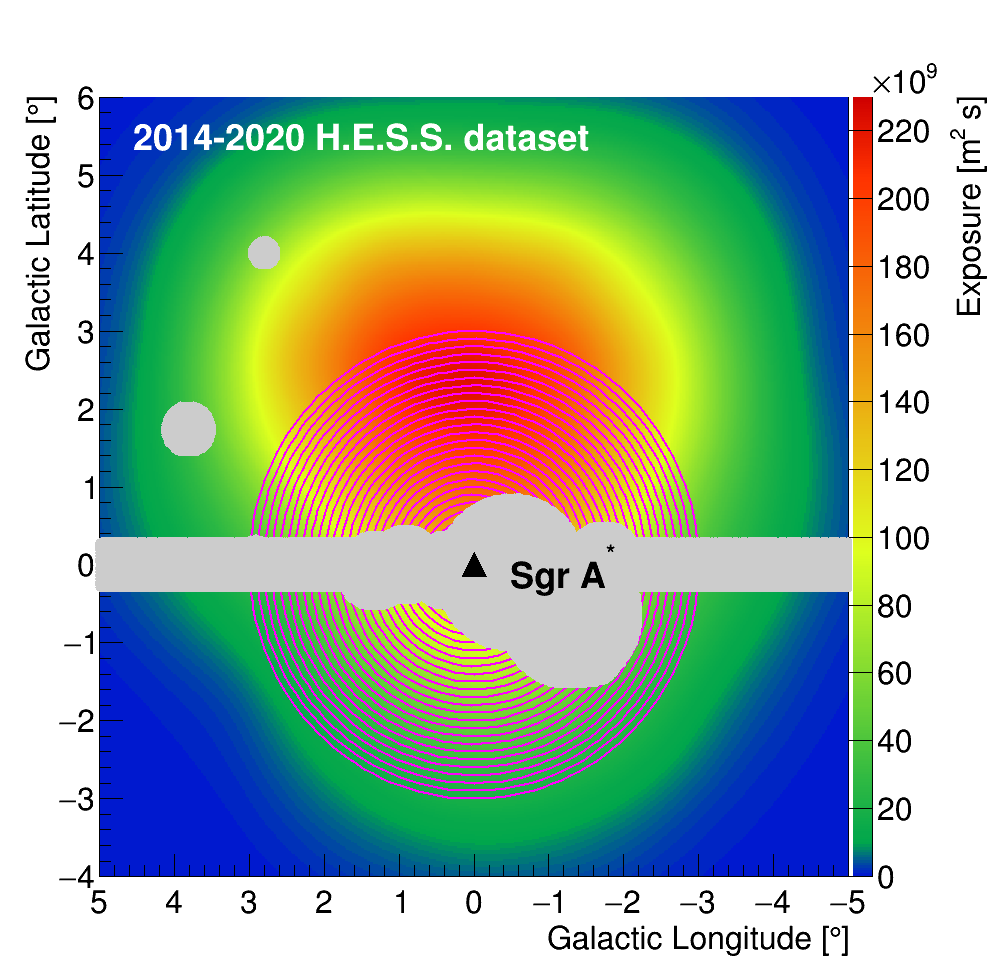}
\includegraphics[width=0.45\textwidth]{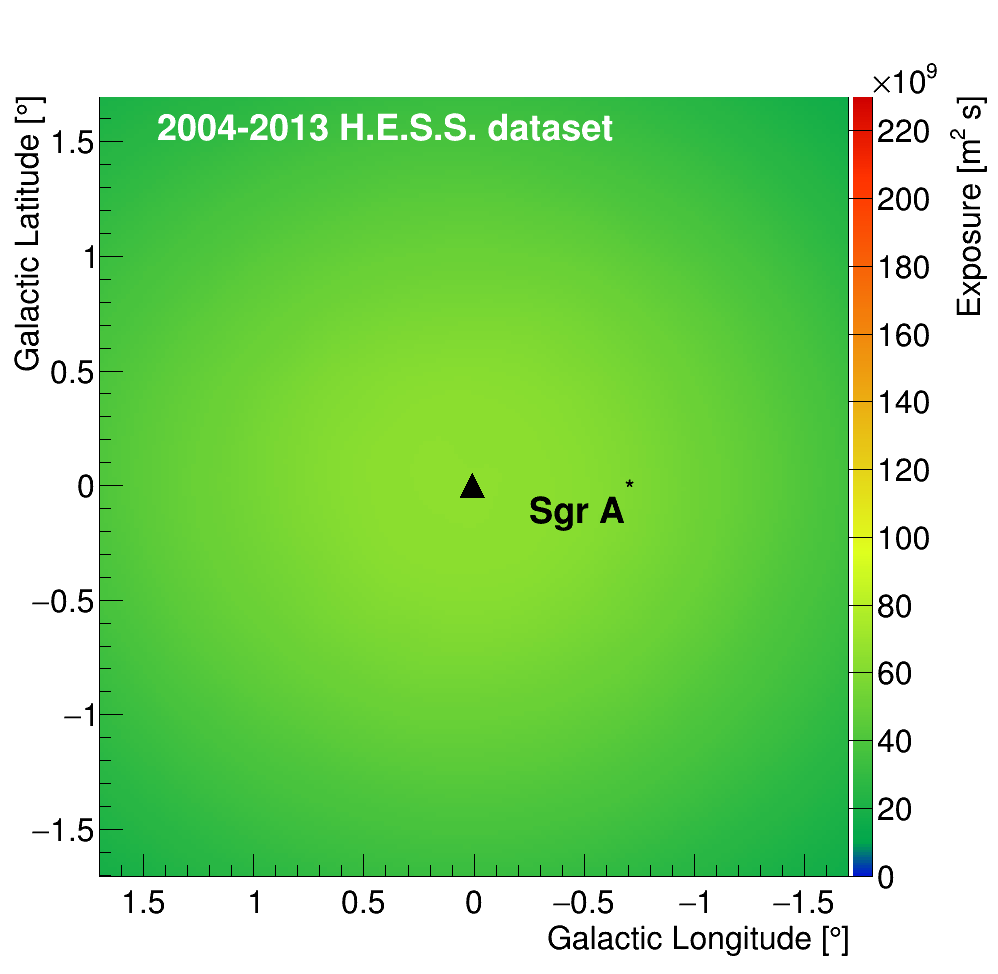}
\includegraphics[width=0.45\textwidth]{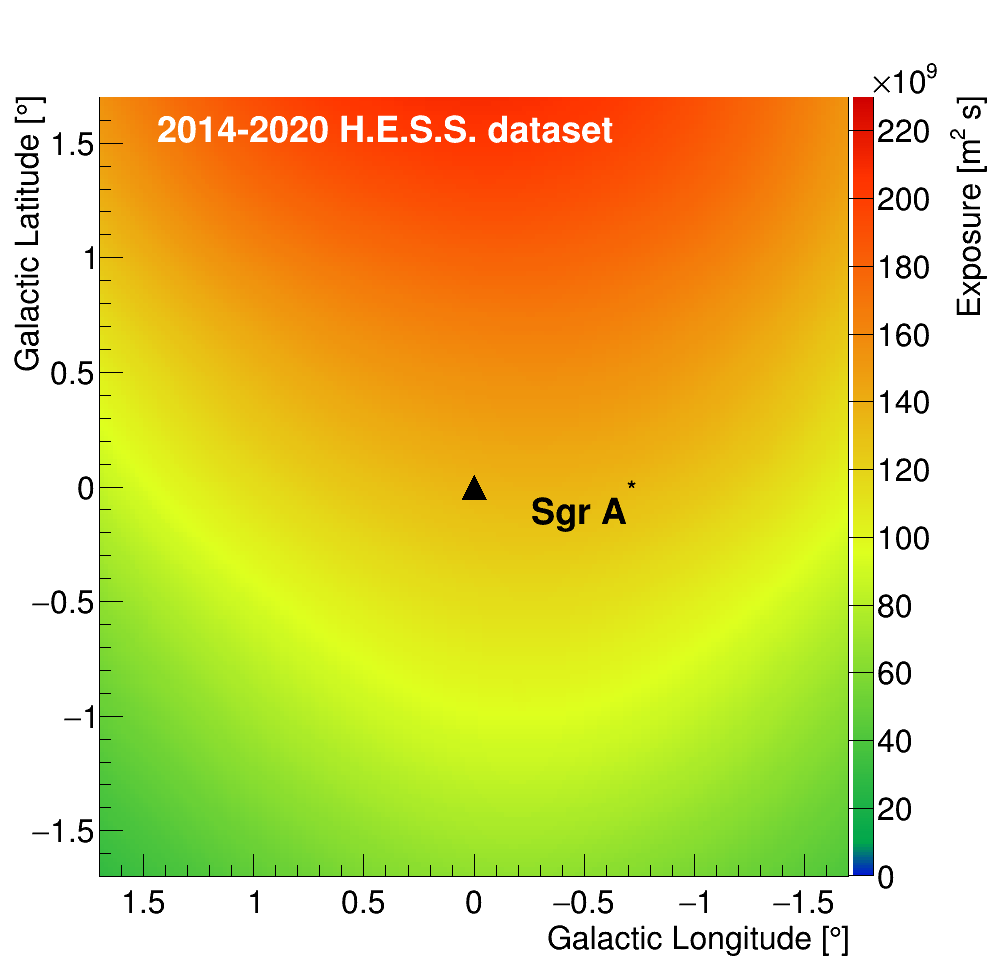}
\caption{Exposure maps (in m$^2$s) in Galactic coordinates from H.E.S.S. observations. The black triangle indicates the position of the supermassive black hole Sgr A*. 
{\it Top-left panel:} Full exposure map obtained with the 2014-2020 dataset used in this work.
The telescope pointing positions of the IGS are displayed as black crosses. 
{\it Top-right panel:} Full exposure map with, overlaid, the regions of interest (ROI, solid magenta lines), defined as 25 annuli centered on the nominal GC position with inner radii from 0.5$^\circ$ to 2.9$^\circ$, and width of 0.1$^\circ$.
The gray-shaded regions show the set of masks used for the exclusion regions in the field of view to avoid astrophysical background contamination from VHE sources in the ROI. 
{\it Bottom-left panel:} Exposure map in Galactic coordinates from H.E.S.S. phase-1 observations. 
{\it Bottom-right panel:} Zoomed view of the exposure map obtained from the 2014-2020 dataset.
}
\label{fig:exposure}
\end{figure}
The top-left panel of Fig.~\ref{fig:exposure} shows the exposure map obtained from the 2014-2020 dataset together with the IGS telescope pointing positions. 
The latter are chosen in order to map the GC region for positive Galactic latitudes. The pointing positions in Galactic coordinates are listed in Tab.~\ref{tab:tabpt}.
The bottom panels of Fig.~\ref{fig:exposure} 
show the exposure map from the H.E.S.S. phase-1 observations used in Refs.~\cite{Abdallah:2016ygi,Abdallah:2018qtu} and a zoomed view of the exposure map obtained from the 2014-2020 dataset, respectively.

After data quality selection~\cite{Aharonian:2006pe} the data set used in the present study amounts to a total of 546 hours. The data are analyzed in stereo mode, {\it i.e.}, requiring at least two telescopes of the array to trigger the same shower event,  with  a  semi-analytical  shower  model~\cite{2009APh32231D}  where
the best event reconstruction
between an array configuration with only the four 12-m diameter telescopes  and one with the five telescopes is chosen.
\begin{table}[h]
\centering
\begin{tabular}{c|c|c|c|c|c|c|c|c|c|c|c|c|c|c}
\hline
\hline
Pointing position name & 1-4 & 1-5 & 1-6 & 1-7 & 1-8 & 1-9 & 2-5 & 2-6 & 2-7 & 2-8 & 3-5 & 3-6 & 3-7 & 3-8 \\
\hline
Gal. long. [deg.] & -3.0 & -1.8 & -0.6 & 0.6 & 1.8 & 3.0 & -1.8 & -0.6 & 0.6 & 1.8 & -1.8 & -0.6 & 0.6 & 1.8 \\
\hline
Gal. lat. [deg.] & 0.8 & 0.8 & 0.8 & 0.8 & 0.8 & 0.8 & 2.0 & 2.0 & 2.0 & 2.0 & 3.2 & 3.2 & 3.2 & 3.2 \\
\hline
\hline
\end{tabular}
\caption{Pointing positions in Galactic coordinates for the 2016-2020 IGS observations. The first row gives the names of the pointing positions, which were chosen sequentially during the years. The second and third rows give the Galactic longitudes and latitudes of the pointing positions.  \label{tab:tabpt}}
\end{table}

\section{Dark matter halo profiles for the Milky Way and J-factor computation in the regions of interest}
\label{sec:jfactor}
\begin{figure}[!ht]
\centering
\includegraphics[width=0.4\textwidth]{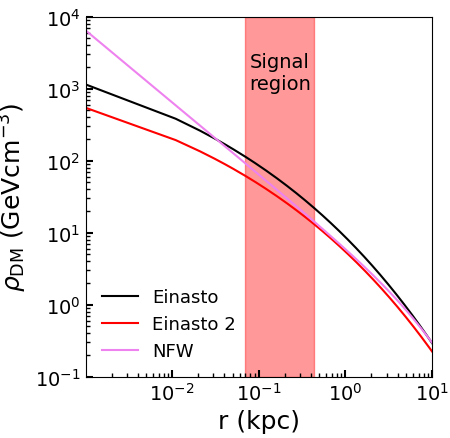}
\caption{Dark matter density profiles $\rho_{\rm DM}$ versus distance $r$ from the Galactic Center. The Einasto and NFW profile parametrizations considered here follow Ref.~\cite{Abdallah:2016ygi}. An alternative parametrization of the Einasto profile~\cite{Cirelli:2010xx} is also used and referred as to "Einasto 2". The red-shaded area corresponds to the signal region where the DM annihilation signal is searched.}
\label{fig:DMdensityprofiles}
\end{figure}

The DM distribution in the Milky Way is assumed to follow a cuspy distribution for which the  Einasto and Navarro-Frenk-White (NFW) profiles are typical parametrizations. They are expressed as:
\begin{equation}
\label{eq:profiles}
\rho_{\rm E}(r) = \rho_{\rm s}  \exp \left[-\frac{2}{\alpha_{\rm s}}\left(\Big(\frac{r}{r_{\rm s}}\Big)^{\alpha_{\rm s} }-1\right)\right]\\
\quad {\rm and} \quad \rho_{\rm NFW}(r) = \rho_{\rm s}\left(\frac{r}{r_{\rm s}}\Big(1+\frac{r}{r_{\rm s}}\Big)^2\right)^{-1}  \ , 
\end{equation}
respectively, assuming a DM density at the Solar position of $\rho_{\odot} = 0.39\ \rm GeV cm^{-3}$~\cite{Catena:2009mf}. $\rho_{\rm s}$ and $r_{\rm s}$ are the scale density and scale radius, respectively. Table~\ref{tab:tab1} provides the parameters of the Einasto and NFW profiles used here, as well as an alternative parameter set for the Einasto profile. 
\begin{table}[h]
\centering
\begin{tabular}{c|c|c|c}
\hline
\hline
Profiles & Einasto & NFW & Einasto 2~\cite{Cirelli:2010xx}  \\
\hline
$\rho_{\rm s}$ (GeVcm$^{-3}$) & 0.079 & 0.307 &  0.033 \\
$r_{\rm s}$ (kpc) 		        & 20.0   & 21.0   & 28.4  \\
$\alpha_{\rm s}$                        & 0.17   &   /       &  0.17  \\
\hline
\hline
\end{tabular}
\caption{Parameters of the cuspy profiles used for the DM distribution. The Einasto and NFW profiles considered here follow Ref.~\cite{Abdallah:2016ygi}. An alternative normalization of the Einasto profile~\cite{Cirelli:2010xx} is also used and referred as to "Einasto 2". 
\label{tab:tab1}}
\end{table}
The DM density profiles for the parametrizations given in Tab.~\ref{tab:tab1} are plotted in Fig.~\ref{fig:DMdensityprofiles}.
\begin{table}[ht!]
\footnotesize
\centering
\begin{tabular}{c|c|c|c|c|c|c}
\hline
\hline
$i^{\rm th}$ ROI & Inner radius & Outer radius & Solid angle $\Delta\Omega$&\multicolumn{3}{c}{$J$-factor $J(\Delta\Omega)$}  \\
 &  [deg.] &  [deg.] & [10$^{-4}$ sr] &\multicolumn{3}{c}{[10$^{20}$ GeV$^2$cm$^{-5}$]}  \\
\hline
& & & & Einasto &  NFW & Einasto~\cite{Cirelli:2010xx} \\
\hline
1 & 0.5 & 0.6 & 1.05 & 9.5 & 4.9 & 3.0 \\ 
2 & 0.6 & 0.7 & 1.24 & 9.8 & 4.9 & 3.2 \\ 
3 & 0.7 & 0.8 & 1.44 & 10.1  & 4.9 & 3.3 \\
4 & 0.8 & 0.9 & 1.63 & 10.2 & 4.8 & 3.4 \\
5 & 0.9 & 1.0 & 1.82 & 10.3 & 4.8 & 3.5 \\
6 & 1.0 & 1.1 & 2.01 & 10.4 & 4.8 & 3.5 \\
7 & 1.1 & 1.2 & 2.20 & 10.5 & 4.7 & 3.6 \\
8 & 1.2 & 1.3 & 2.39 & 10.5 & 4.7 & 3.6 \\
9 & 1.3 & 1.4 & 2.58 & 10.5 & 4.7 & 3.6 \\
10 & 1.4 & 1.5 & 2.77 & 10.5 & 4.6 & 3.7 \\
11 & 1.5 & 1.6 & 2.97 & 10.4 & 4.6 & 3.7 \\
12 & 1.6 & 1.7 & 3.16 & 10.4 & 4.6 & 3.7 \\
13 & 1.7 & 1.8 & 3.35 & 10.3 & 4.5 & 3.7 \\
14 & 1.8 & 1.9 & 3.54 & 10.3 & 4.5 & 3.7 \\
15 & 1.9 & 2.0 & 3.73 & 10.2 & 4.5 & 3.7 \\
16 & 2.0 & 2.1 & 3.92 & 10.2 & 4.5 & 3.7 \\
17 & 2.1 & 2.2 & 4.11 & 10.1 & 4.4 & 3.7 \\
18 & 2.2 & 2.3 & 4.31 & 10.0 & 4.4 & 3.7 \\
19 & 2.3 & 2.4 & 4.50 & 9.9 & 4.4 & 3.7 \\
20 & 2.4 & 2.5 & 4.69 & 9.9 & 4.3 & 3.6 \\
21 & 2.5 & 2.6 & 4.88 & 9.8 & 4.3 & 3.6 \\
22 & 2.6 & 2.7 & 5.07 & 9.7 & 4.3 & 3.6 \\
23 & 2.7 & 2.8 & 5.26 & 9.6 & 4.3 & 3.6 \\
24 & 2.8 & 2.9 & 5.45 & 9.5 & 4.3 & 3.6 \\
25 & 2.9 & 3.0 & 5.64 & 9.5 & 4.2 & 3.6 \\
\hline
\hline
\end{tabular}
\caption{J-factor values in units of GeV$^2$cm$^{-5}$ in each of the 25 ROI considered in this work. The first four columns give the ROI number, the inner radius, the outer radius, and the size in solid angle for each RoI. The fifth column provides the total J-factor values in the ROI, {\it i.e.}, computed without applying the masks on the excluded regions, for the Einasto profile considered in this work together with the values obtained for an NFW profile~\cite{Abdallah:2016ygi} and an alternative normalization of the Einasto profile~\cite{Cirelli:2010xx} in sixth and seventh columns, respectively. \label{tab:jfactors}}
\end{table}

The total J-factor values are computed for each region of interest (ROI).
Table~\ref{tab:jfactors} provides the inner and outer radii for each ROI, their solid angle and the corresponding total J-factor values for Einasto and NFW profiles. 
The J-factor map for the Einasto profile is shown in left panel of Fig.~\ref{fig:reflected_background}.

\section{Definition of the regions of interest, measurement of the residual background, photons statistics and fluxes}
The observations are performed with all the telescopes pointed in the same direction in the sky and the data taking requires that at least two telescopes are triggered by the same air shower event. Given the field of view of the H.E.S.S. instrument and the pointing positions, 
a significant event statistics is obtained up to about 6$^\circ$ above the Galactic plane. We search for a DM annihilation signal in a disk centered at the Galactic Centre with radius of 3$^\circ$. In order to benefit from the different 
spatial morphology of the searched signal with respect to the background, the disk is further divided into 25 ROIs defined as rings of inner radii from 0.5$^\circ$ to 2.9$^\circ$. The width of each ring is 0.1$^\circ$. Each ROI is defined as the ON region. On the top-right panel of Fig.~\ref{fig:exposure} the 25 ROI are overlaid.

\begin{figure}[!ht]
\centering
\includegraphics[width=0.32\textwidth]{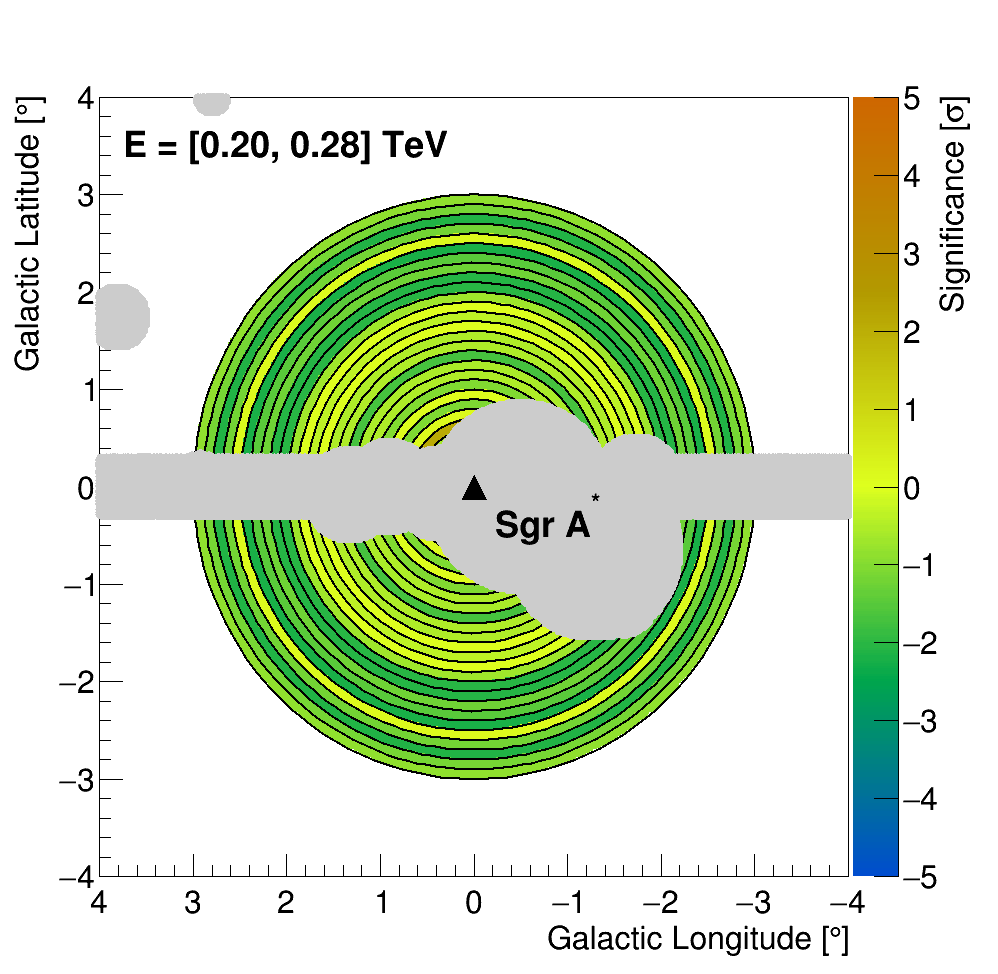}
\includegraphics[width=0.32\textwidth]{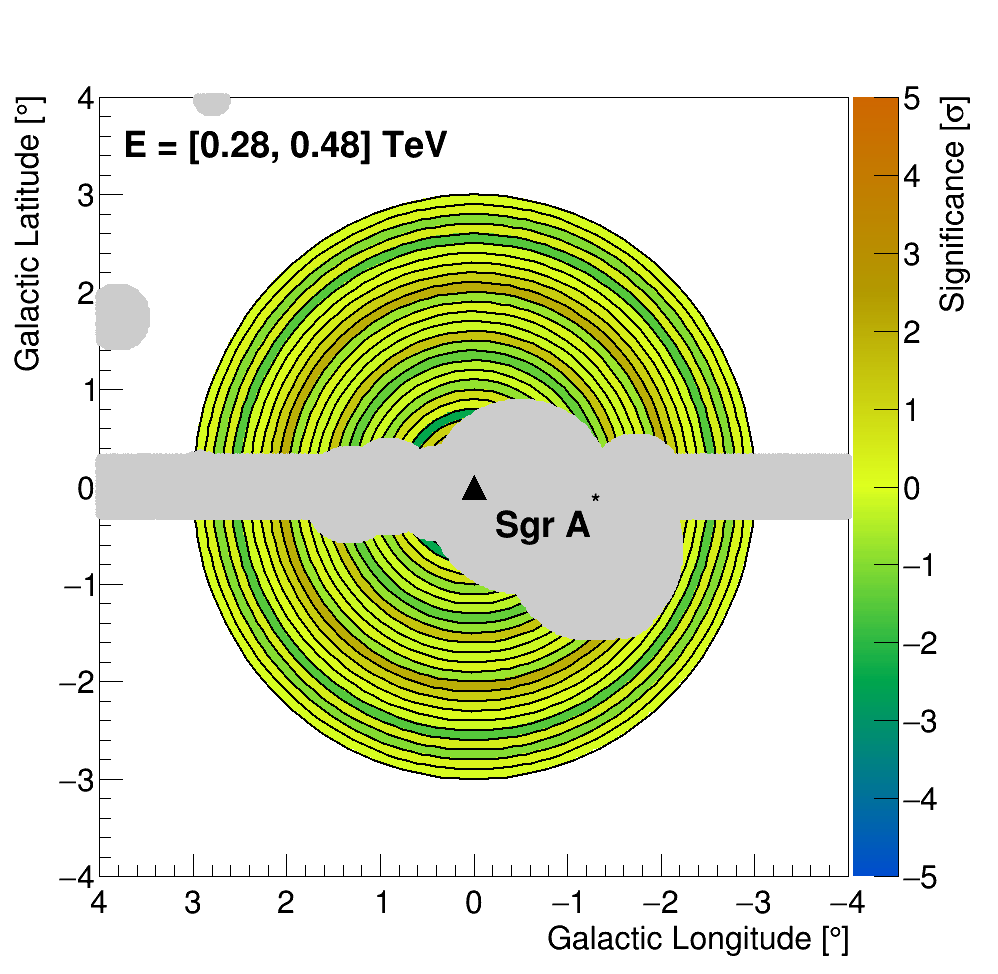}
\includegraphics[width=0.32\textwidth]{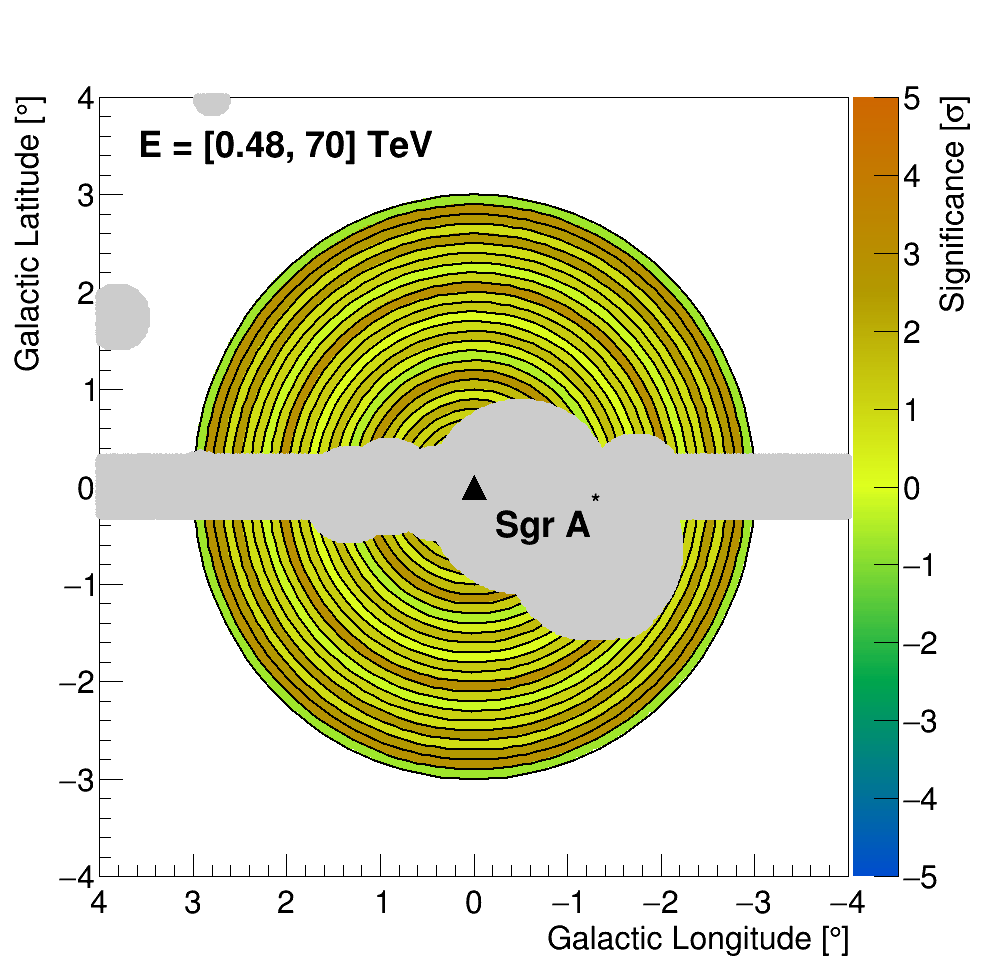}
\caption{Significance map of the residuals in Galactic coordinates in three energy bands. 
The grey-shaded region corresponds to the set of masks used in this analysis to avoid astrophysical background contamination from the VHE sources in the ROIs.  The black triangle shows the position of the supermassive black hole Sagittarius A*.
}
\label{fig:dists}
\end{figure}

The GC region is a very rich and complex environment at VHE energies. It harbors numerous cosmic-ray sources producing VHE gamma-rays including the supermassive black hole Sagittarius A*, pulsar wind nebulae and supernova remnants. A set of conservative masks is used in order to avoid VHE gamma-ray contamination both in the signal and in the background regions. For each pointing position 
and each run of the dataset, the background is measured simultaneously in the same field of view as used for signal search in an OFF region taken symmetrically to the ON region with respect to the pointing position, as described in Refs.~\cite{Abramowski:2011hc,Abdallah:2016ygi}. Therefore, the expected background in the ON region is determined from a OFF region measurement performed under the same observational and instrumental conditions. The regions of the sky with VHE gamma-ray sources are excluded for both the ON and OFF measurements, providing them with the same solid angle size. This background measurement technique is carried out on a run-by-run basis and provides an accurate determination of the  residual background.
For a given ROI and a given pointing position of a run, the OFF events are measured. For a different run with the same ROI and same pointing position, the OFF events fall in the same spatial region but they are totally independent from the previous observation run. 
For each ROI, the photon statistics in the ON and OFF regions as well as the corresponding excess significance are provided in Tab.~\ref{tab:statistics}. 
\begin{table}[!hb]
\centering
\begin{tabular} {c|c|c|c|c|c|c|c|c|c|c|c|c}
\hline
\hline
$i^{\rm th}$ ROI & 1 & 2 & 3 & 4 & 5 & 6 & 7 & 8 & 9 & 10 & 11 & 12 
\\
\hline
\hline
N$_{\rm ON}$ 
& 326 & 1830 & 3029 & 4736 & 6793 & 9144 & 12036 & 15201 & 16830 & 19530 &	23549 &	25585 
\\
\hline
N$_{\rm OFF}$ 
& 298 & 1674 & 3087 &	4665 & 6699 & 9164 & 11899 & 15177 & 17242 & 19721 & 23270 & 25568 
\\
\hline
S($\sigma$) 
& 1.1 & 2.6 & -0.7 & 0.7 & 0.8 & -0.2 & 0.9 & 0.1 & -2.2 & -0.9 & 1.3 & 0.1 
\\
\hline
\hline
13 & 14 & 15 & 16 & 17 & 18 & 19 & 20 & 21 & 22 & 23 & 24 & 25\\
\hline
27571 & 29875 & 32328 & 35094 & 37292 & 39957 & 42540 & 42460 & 42282 & 42317 & 42653 & 43188 & 42879
\\
\hline
27673 & 29945 & 32518 & 34774 & 37502 & 40159 & 42775 & 42939 & 42415 & 42509 & 42896 & 43011 & 43373
\\
\hline
-0.4 & -0.3 & -0.8 & 1.2 & -0.8 & -0.7 & -0.8 & -1.6 & -0.5 & -0.7 & -0.8 & 0.6 & -1.7
\\
\hline
\hline
\end{tabular} 
\caption{Photon statistics in the ON and OFF regions, respectively, together with the corresponding excess significance in each of the 25 ROIs considered in this work. The first row gives the ROI number. The second and third columns give the photon statistics in the ON and OFF regions, respectively. The fourth row gives the excess significance. 
\label{tab:statistics}}
\end{table}
Fig.~\ref{fig:dists} shows the significance maps of the residuals for three energy bands. The energy bands are chosen such that comparable photon statistics is found in each one. 
 In the high energy band, there is an overall significant gamma-ray excess, with the formal total significance of 5.7$\sigma$ which corresponds to the p-value of 1.1x10$^{-8}$. This suggests the presence of an unaccounted additional background signal. 
However, the spatial and spectral shapes of this signal 
are not compatible with the expected ones from the DM signal.
Nevertheless, this additional contribution is considered as part of the measured excess events in our analysis, which makes the constraints on $\langle \sigma v \rangle$ conservative since the observed excess is positive.

\begin{figure}[!ht]
\centering
\includegraphics[width=0.44\textwidth]{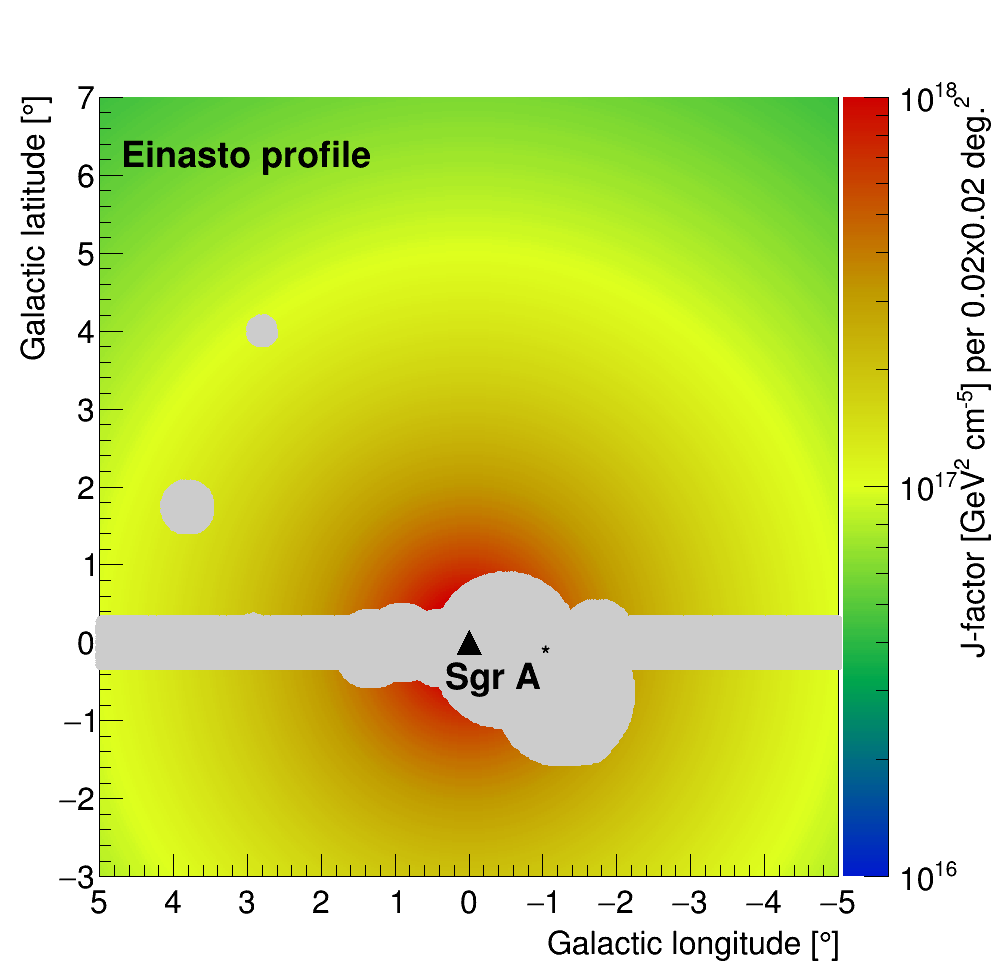}
\hspace{1cm}
\includegraphics[width=0.44\textwidth]{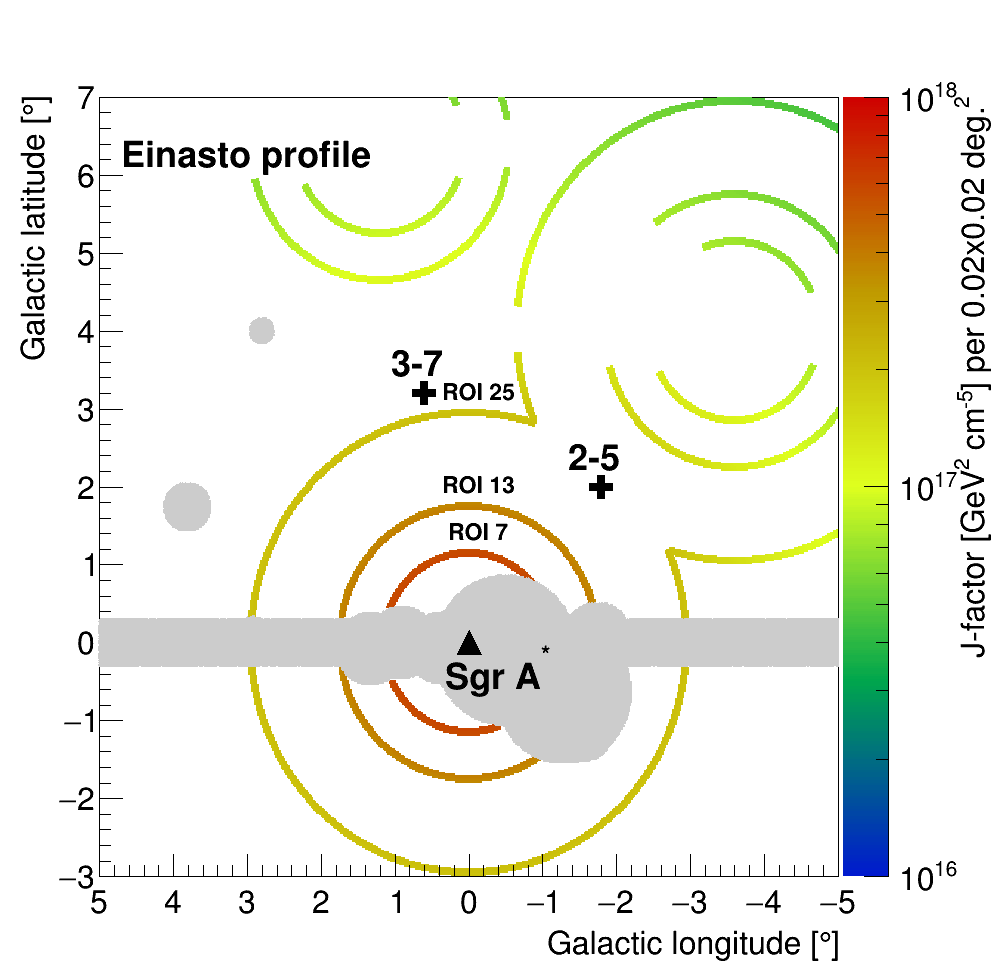}
\caption{
{\it Left panel:} J-factor map for the Einasto profile in Galactic coordinates. The J-factor values are integrated in pixels of 0.02$^\circ \times$0.02$^\circ$ size. The grey-shaded region corresponds to the set of masks used in this analysis to avoid astrophysical background contamination from the VHE sources in the ROIs. The black triangle shows the position of the supermassive black hole Sagittarius A*.
{\it Rigth panel:} Background determination method in Galactic coordinates. Two IGS pointing positions are marked with black crosses. J-factor values are displayed for the ROI 7 and 13, respectively, together with those obtained in the corresponding OFF regions.  In addition, the J-factor values for ROI 25 and its corresponding OFF region with respect to the pointing position 2-5 are shown. The masked regions are excluded similarly in the ON and OFF regions such that these regions keep the same solid angle size and acceptance. The black triangle shows the position of the supermassive black hole Sagittarius A*. 
}
\label{fig:reflected_background}
\end{figure}

The right panel of Fig.~\ref{fig:reflected_background} shows an example of the background measurement for the ROI 7 and 13 and the pointing positions (black crosses) 2-5 ($l$ = -1.8$^\circ$, $b$ = 2.0$^\circ$) and 3-7 ($l$ = 0.8$^\circ$, 
$b$ = 3.2$^\circ$), respectively. The background measurement for ROI 25 and pointing position 2-5 is also shown. The set of masks used in this analysis is shown as a grey shaded area. Masks include the Galactic plane  between $\pm$0.3$^{\circ}$ as well as the diffuse emission region around the GC~\cite{Abramowski:2016mir}, sources from Ref.~\cite{H.E.S.S.:2018zkf}, and all VHE gamma-ray sources in the field of view. The exclusion regions are removed similarly in the ON and OFF regions such that they keep the same solid angle and acceptance.
The color scale indicates the value of the J-factor computed for the Einasto profile in the pixel size of 0.02$^\circ \times$0.02$^\circ$.  
The ratio between the J-factor values in ON and OFF regions for ROI 13 with respect to pointing positions 3-7 and 2-5 are 5 and 4, respectively, which maintains a significant expected DM excess signal in the ON region with respect to the OFF region.

\begin{figure}[!ht]
\centering
\includegraphics[width=\textwidth]{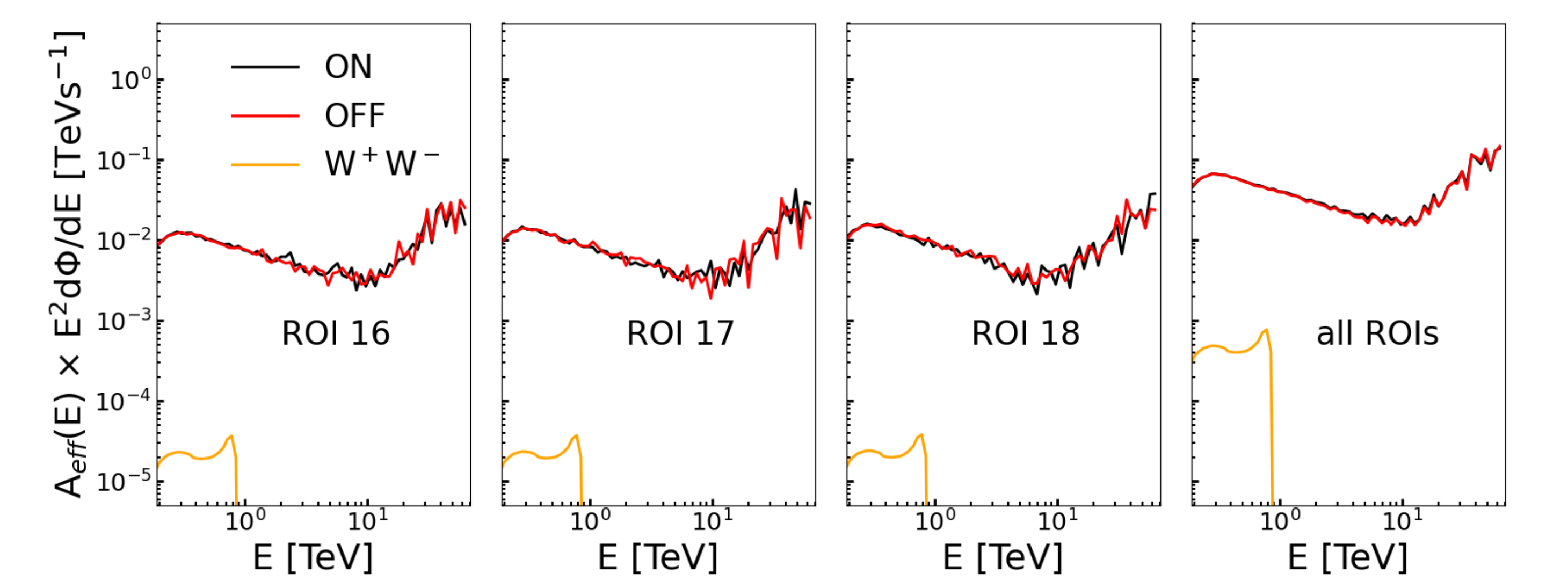}
\caption{Energy-differential spectra expected for the self-annihilation of DM with mass $m_{\rm DM}$ = 0.98 TeV and $\langle \sigma v \rangle = 3.8 \times 10^{-26}$ cm$^3$s$^{-1}$ in the $W^+W^-$ annihilation channel 
 multiplied by $E^2$ and
convolved with the H.E.S.S. response (orange line) for individual ROIs as well as for the combination of all ROIs. 
$A_{\rm eff}(E)$ stands for the energy-dependent acceptance of the instrument.
Also plotted are the corresponding ON (black line) and OFF (red line) energy-differential spectra.
}

\label{fig:acceptancespectrum}
\end{figure}

Figure~\ref{fig:acceptancespectrum} shows the energy-differential annihilation spectrum in the $W^+W^-$ channel convolved with the H.E.S.S. acceptance and energy resolution expected for the self-annihilation of DM with mass $m_{\rm DM}$ = 0.98 TeV and $\langle \sigma v \rangle = 3.8 \times 10^{-26}$ cm$^3$s$^{-1}$ for individual ROIs as well as for the combination of all ROIs.
Overlaid are the corresponding ON and OFF energy-differential spectra convolved with the H.E.S.S. energy-dependent acceptance ($A_{\rm eff}$) and energy resolution. 

\begin{figure}[!ht]
\includegraphics[width=\textwidth]{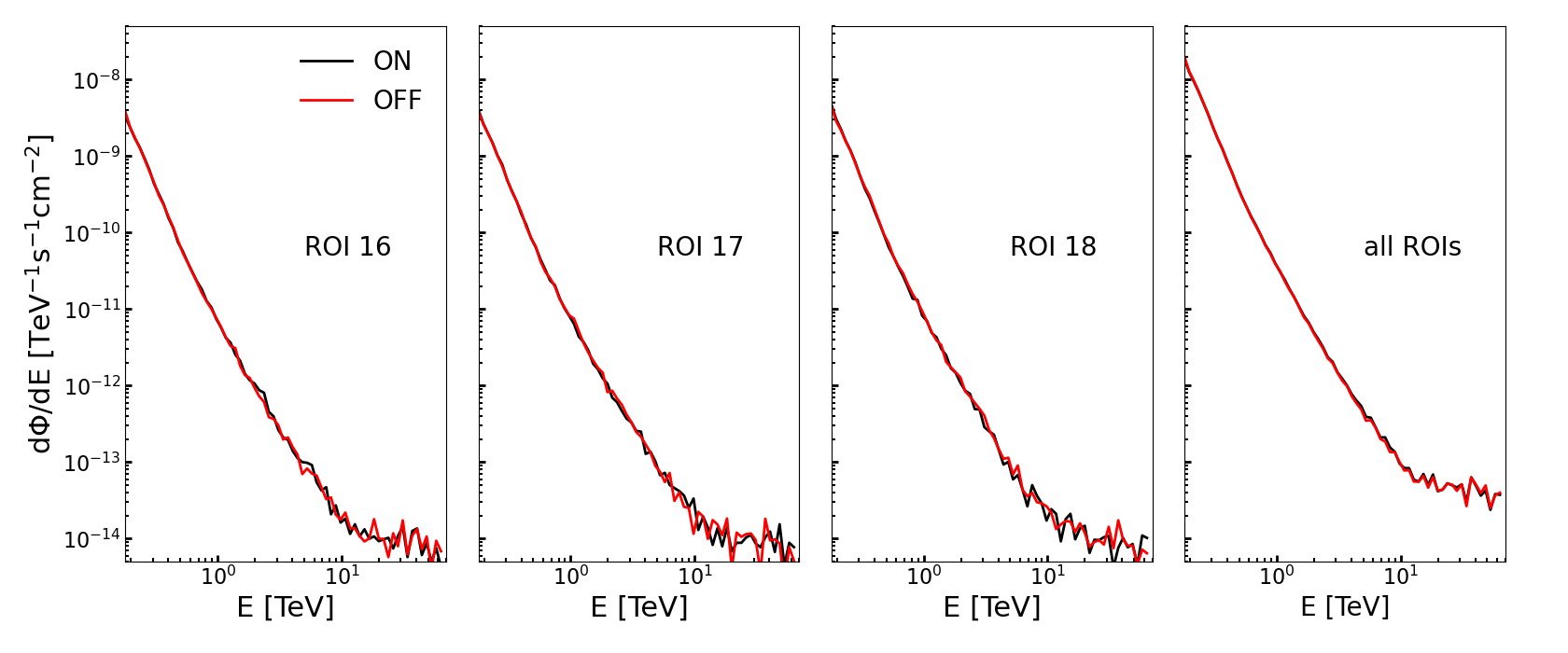}
\caption{The energy-differential flux for ON (black line) and OFF (red line) regions for individual ROIs are shown in the first three panels for ROI 16, 17 and 18, respectively. The right panel shows ON and OFF energy-differential fluxes for the combination of all ROIs.}
\label{fig:flux}
\end{figure}

\begin{figure}[!ht]
\includegraphics[width=0.8\textwidth]{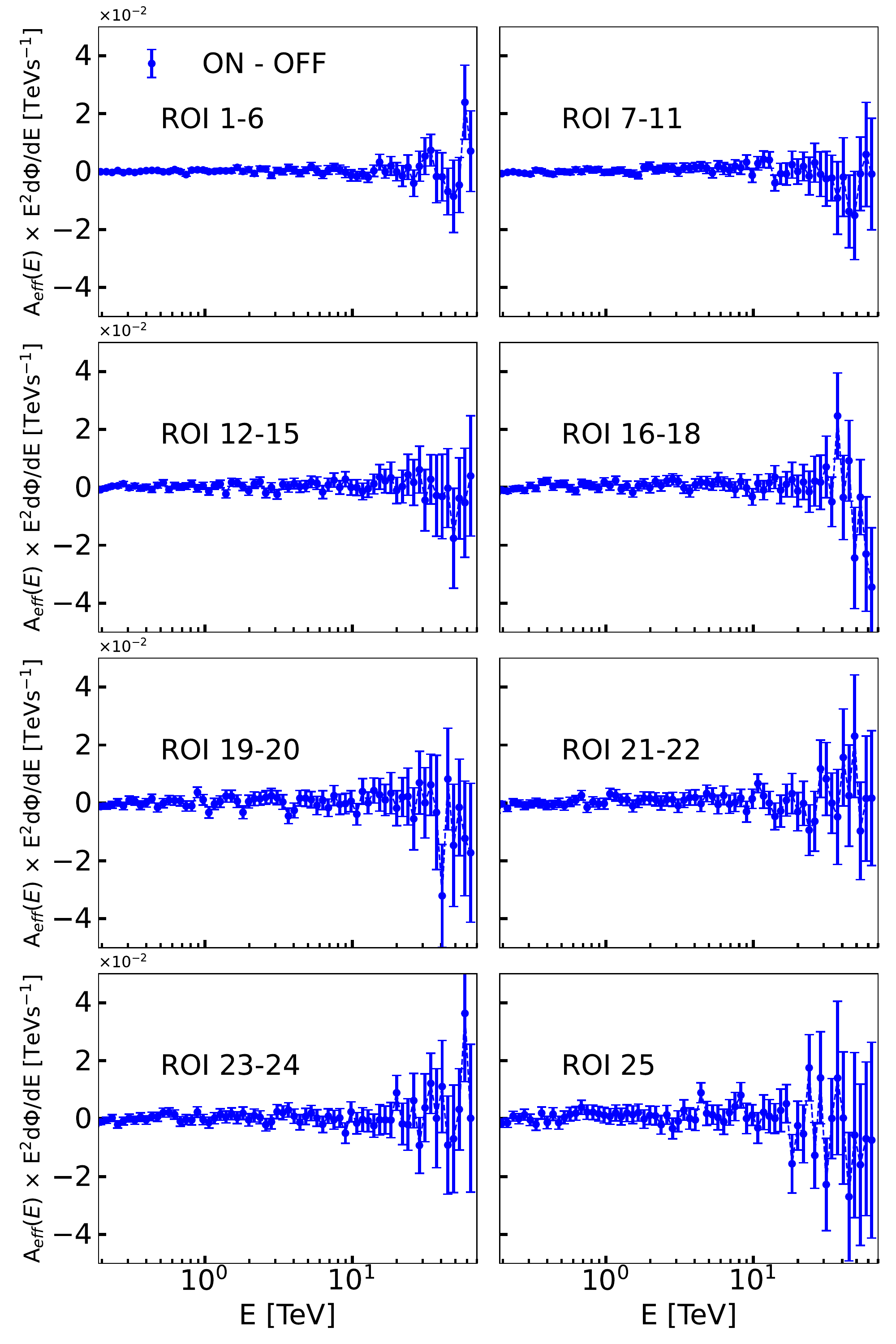}
\caption{Background-subtracted energy-differential fluxmultiplied by $E^2$ and convolved with the H.E.S.S. response, versus energy for different combinations of ROIs. 
$A_{\rm eff}(E)$ stands for the energy-dependent acceptance of the instrument.
1$\sigma$ error bars are shown.From the top to the bottom, the following ROI combinations are shown: from ROI 1 up to 6, from 7 up to 11, from 12 up to 15, from 16 up to 18, 19 and 20, 21 and 22, 23 and 24, and 25.}
\label{fig:residualflux}
\end{figure}

In Fig.~\ref{fig:flux} are plotted the 
energy-differential flux for ON and OFF regions  for individual ROIs  as well as for the combination of all ROIs.
The steep spectrum of the residual background is mainly due to the dominant contribution of  misidentified cosmic-rays.
Fig.~\ref{fig:residualflux} shows the background-subtracted energy-differential flux, convolved with the H.E.S.S. response, for different combinations of the ROIs as explained in the caption. 

\section{Statistical method for upper limit determination}
The statistical analysis method makes use of a log-likelihood ratio test statistic (TS) to test the DM signal hypothesis against the data assuming a positive searched signal, {\it i.e.}, $\langle  \sigma v \rangle >$ 0.  Following 
Ref.~\cite{2011EPJC711554C}, the TS is defined as: 
\begin{equation}
TS = 
\begin{cases}
- 2\, \rm ln \frac{\mathcal{L}(N^{\rm S}(\langle \sigma v \rangle), \widehat{\widehat {N^{\rm B}}}(\langle \sigma v \rangle)
)}{\mathcal{L}(0,\widehat{\widehat{N^{\rm B}(0)}}
)} \,  & N^{\rm S}(\widehat{\langle \sigma v \rangle}) < 0\\
- 2\, \rm ln \frac{\mathcal{L}(N^{\rm S}(\langle \sigma v \rangle), \widehat{\widehat {N^{\rm B}}}(\langle \sigma v \rangle)
)}{\mathcal{L}(N^{\rm S}(\widehat{\langle \sigma v \rangle}),\widehat{N^{\rm B}}
)} \,  & 0\leq N^{\rm S}(\widehat{\langle \sigma v \rangle}) \leq N^{\rm S}(\langle \sigma v \rangle)\\
0 & N^{\rm S}(\widehat{\langle \sigma v \rangle}) > N^{\rm S}(\langle \sigma v \rangle) \, .
\end{cases}
\label{eq:TS}
\end{equation}
$N^{\rm S}$ is obtained summing $N^{\rm S}_{\rm k}$ over all the runs $k$, where $N^{\rm S}_{\rm k}$  corresponds to the number of  gamma rays expected from DM annihilation for the observational run $k$. 
From Majorana DM particles of mass $m_{\rm DM}$
self-annihilating with a thermally-averaged annihilation cross section
$\langle \sigma v \rangle$ in the channels $f$ with differential spectra $dN^f_{\gamma}/dE_{\gamma}$ of branching ratios $BR_f$,
in a region of solid angle $\Delta\Omega$ with a J-factor $J(\Delta\Omega$),  $N^{\rm S}_{\rm k}$ is given by:
\begin{equation}
 N^{\text{S}}_{\text k}(\langle \sigma v \rangle) =  \frac{\langle \sigma v \rangle J(\Delta\Omega)}{8\pi m_{\rm DM}^2} T_{\rm{obs},k} \int_{E_{\rm th}}^{m_{\rm DM}} \int^{\infty}_{0}   \sum_f BR_f \frac{dN^f_{\gamma}}{dE_{\gamma}}(E_{\gamma}) \: R(E_{\gamma}, E'_{\gamma}) \: A_{\rm eff, k}(E_{\gamma}) 
 \:  dE_{\gamma} \: dE'_{\gamma}\, ,
\end{equation}
where the finite energy resolution $R(E_{\gamma}, E'_{\gamma})$ relates the energy detected $E'_{\gamma}$ to the true energy $E_{\gamma}$ of the events,  $A_{\rm eff, k}(E_{\gamma})$ is the energy-dependent acceptance for the run $k$, and $T_{\rm obs ,k}$ is the observation time of the run $k$. The energy-dependent acceptance is computed according to the semi-analytical shower model template technique using standard selection cuts~\cite{2009APh32231D}. For each run, the spatial response of the instrument is encoded in the acceptance term which depends on the angular distance between the reconstructed event position and the pointing position of the run $k$.  Spatial responses of the H.E.S.S. instrument can be found, for instance, in Ref.~\cite{Aharonian:2006pe}. The energy resolution is well described by a Gaussian function of $\sigma/E$ of 10\% above 200~GeV~\cite{2009APh32231D}. 

$\widehat{\widehat {N^{\rm B}_{\rm ij}}}$ is obtained through a conditional maximization by solving $\partial \mathcal{L}/\partial N^{\rm B}_{\rm ij} = 0$. It represents the conditional maximum likelihood estimator of $\mathcal{L}$, {\it i.e.}, the value of $N^{\rm B}_{\rm ij}$ that maximizes $\mathcal{L}$ for the specified $N^{\rm S}_{\rm ij}(\langle \sigma v \rangle)$. $N^{\rm S}_{\rm ij}(\widehat{\langle \sigma v \rangle})$ and $\widehat{N^{\rm B}_{\rm ij}}$ are computed using an unconditional maximization, {\it i.e.}, they represent the maximum likelihood estimators of $\mathcal{L}$.

As no significant VHE gamma-ray excess is found in any of the ROI, the TS enables to derive upper limits on the thermally-averaged velocity-weighted annihilation cross section $\langle  \sigma v \rangle$ for a set of DM masses $m_{\rm DM}$ 
and annihilation spectra. 
95\% C. L. one-sided upper limits are computed via the TS by demanding a TS value of 2.71 assuming that the TS follows a $\chi^2$ distribution, as expected in the high statistics limit, with one degree of freedom.

The analysis makes use of the expected spectral and spatial characteristics of the searched DM signal with respect to residual background. Therefore, the 
total likelihood function is given by the product of Poisson likelihood functions over the spatial and energy bins $\mathcal{L} = \prod_{\rm ij}\mathcal{L}_{\rm ij}$.
The two-body DM annihilation is taking place almost at rest, the DM-induced gamma-ray spectrum in the final state is expected to exhibit a sharp energy cut-off at the DM mass with possible bump-like energy features close to the DM mass. These spectral characteristics provides efficient discrimination against the much smoother power-law like spectrum of the residual background. In addition, the spatial morphology of the expected DM signal follows the spatial J-factor profile, which provides additional sensitivity given the spatially-independent morphology of the residual background.

\section{Expected limit computation}
\label{sec:expectation}
For each mass and each annihilation channel, the 95\% C.L. expected limits are derived
from a set of 300 Poisson realizations of the measured background event distributions.
For each ROI and each run, an independent Poisson realization of the measured background event energy distribution is computed 
for the ON and the OFF regions, respectively. This provides an overall realization of the expected ON and OFF energy count distributions which is obtained by summing the realizations over all the runs of the dataset. For each realization of the overall energy count distribution in the ON and OFF regions, the corresponding value of  $\langle \sigma v \rangle$ is computed according to the test statistics given in Eq.~(\ref{eq:TS}). This procedure is repeated 300 times. 
The mean expected limits, the 68\% and 95\% statistical containment bands are given by the mean, the 1 and 2$\sigma$ standard deviations, obtained by the distribution of the computed limits. The mean expected limits and the containment bands are plotted in Fig.~\ref{fig:results_channels}.

\section{Study of the systematic uncertainties}
The inner few degrees of the Galactic Centre region is a complex environment with numerous sources emitting in the high and very-high-energy gamma-ray regimes. A conservative set of masks is used to exclude all the regions of the sky with VHE emissions and therefore avoid leakage from nearby sources both into the ON and OFF regions. The H.E.S.S. PSF is 0.06° at 68\% containment radius above 200 GeV~\cite{2009APh32231D} improving slightly with energy (see, from instance, Fig. 25 of Ref.~\cite{2009APh32231D}). For pointlike sources, a circular mask of 0.25$^\circ$ radius is used, the remaining signal leaking outside the source mask is much less than 1\%. In the case of the extended source HESS J1745-303, a circular mask of 0.9$^\circ$ radius is used (see the set of used masks shown as a grey-shaded region in the top-right panel of Fig.~\ref{fig:exposure}). The level of the Night Sky Background (NSB) is subject to significant changes due to the presence of bright stars in the field of view, varying from 100 MHz up to 350 MHz photoelectron rate per pixel in the field of view. A dedicated treatment of the NSB is performed in the shower template analysis method as described in Ref.~\cite{2009APh32231D}, where the contribution of the NSB is modelled in every pixel of the camera. This analysis method does not require any further image cleaning to extract the pixels illuminated by the showers. The method used for the background determination implies that the event counts in the OFF region may be measured in regions of the sky with a different level of NSB compared to the ON region. However, as mentioned above, this can be properly handled with the analysis method used here~\cite{2009APh32231D}.

For each run with a given telescope pointing position and each ROI, background events are measured in an OFF region defined as the region symmetric to the ON region with respect to the pointing position. The residual background rate is correlated with the zenith angle of the observation. For a pointing position at a given zenith angle, a gradient in the residual background rate is expected across the telescope field of view.  
A difference in the zenith angles of the events is obtained between the ON and OFF regions. For the considered ROIs and pointing positions of the IGS, the difference of the means of the distributions of the ON and OFF event zenith angles is up to 1$^\circ$ depending on the zenith angle of the observational run.
The gradient of the gamma-ray-like rate in the FoV is taken into account on a run-by-run basis: 
for each run, the gamma-ray-like rate is renormalized according to the difference of the zenith angle means of the ON and OFF distributions, $\widehat{\theta_z^{ON}}$ and $\widehat{\theta_z^{OFF}}$, respectively, such as $N_{\rm OFF, renorm} = 1.01 \times N_{\rm OFF}  \times(\widehat{\theta_z^{ON}} - \widehat{\theta_z^{OFF}})/1^\circ$.
The mean zenith angle of the observational runs of the dataset is 18$^\circ$. For a run taken at this zenith angle, the difference of the zenith angle means of the ON and OFF distributions is up to 0.5$^\circ$.
To account for the typical width of 1$^\circ$ of the zenith angle distribution,
a systematic uncertainty of 1\% for the normalisation of the measured energy count distributions is used.
The systematic uncertainty derived on the normalisation 
of the energy count distributions deteriorates the mean expected limits from 8\% to 18\% depending on the DM particle mass.

A systematic uncertainty may arise from the assumption of azimuthal symmetry in the field of view. For a given pointing position, the number of counts is computed as a function of the angle. No significant effect is observed beyond the expected 1\%-per-degree gradient in the FoV.  

In the present dataset, the systematic uncertainty on the energy scale of the energy count distributions is 10\%. This systematic uncertainty affects similarly the energy scale of the measured and expected energy count distributions. 
Therefore, the 10\% shift of the energy scale leads to an overall shift of the limits curves along the DM mass axis by 10\%. This systematic uncertainty is not included in the limits.

\section{Upper limits in several annihilation channels}
\label{sec:channels}
\begin{figure*}[!hb]
\centering
\includegraphics[width=0.45\textwidth]{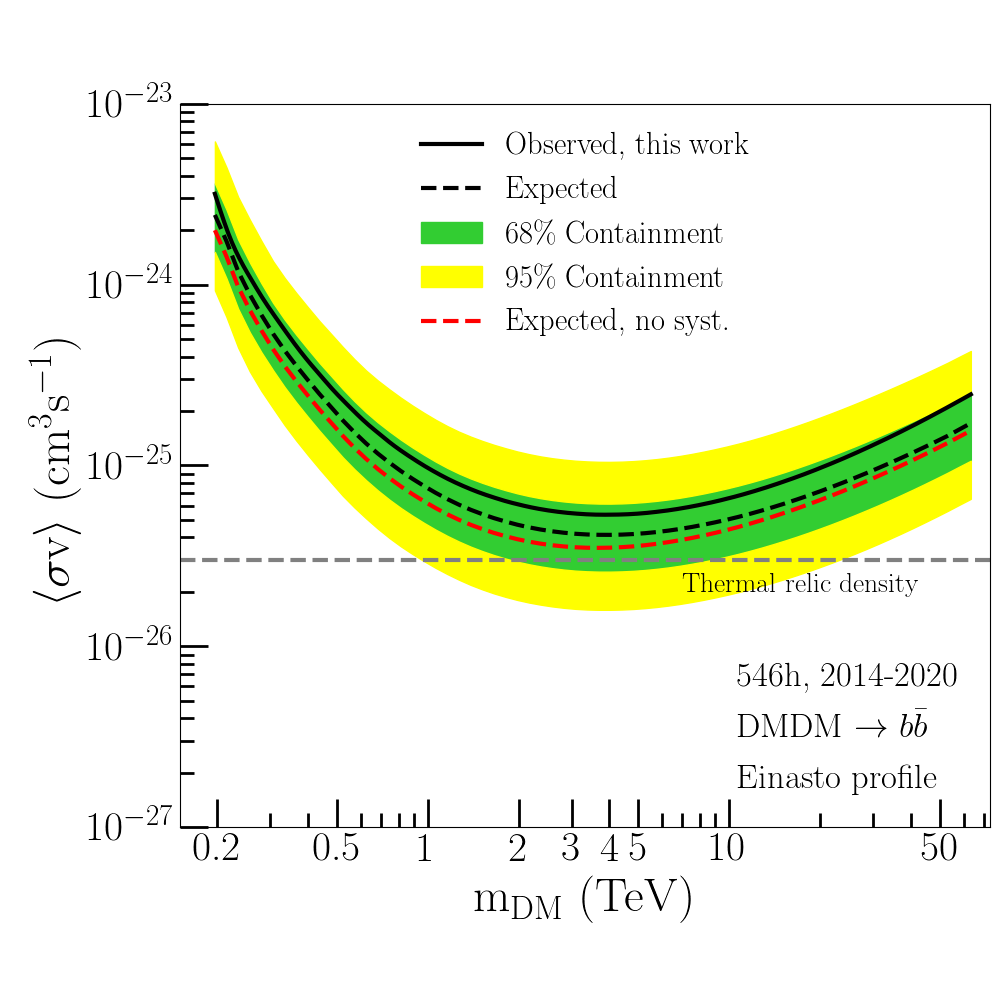}
\vspace{-0.7cm}
\includegraphics[width=0.45\textwidth]{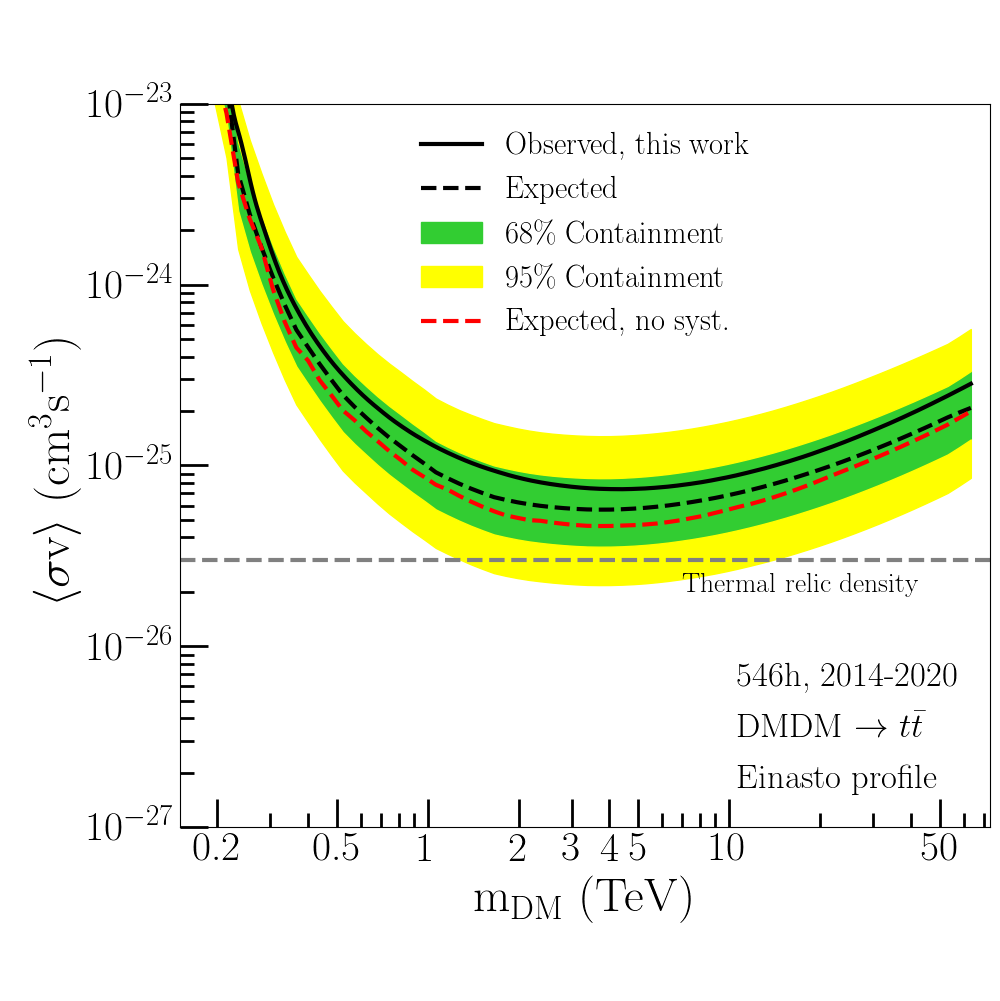}
\vspace{-0.7cm}
\includegraphics[width=0.45\textwidth]{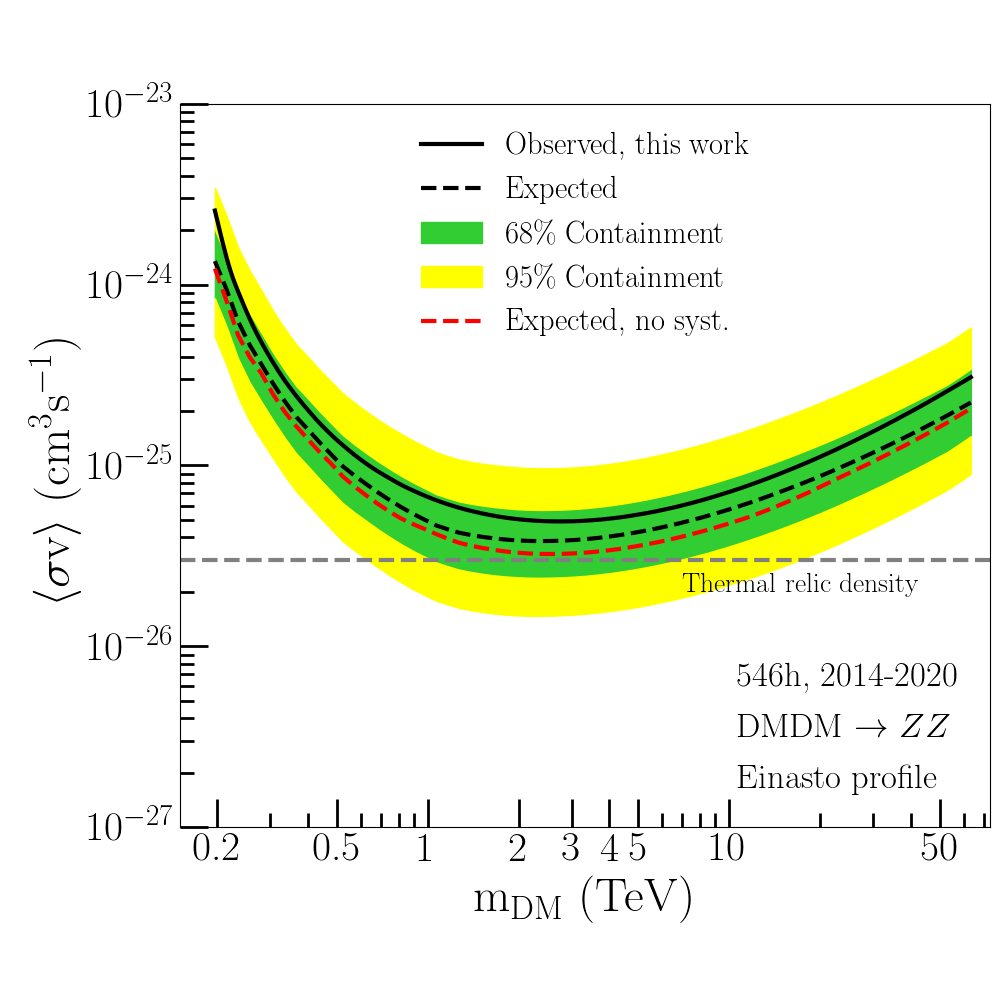}
\includegraphics[width=0.45\textwidth]{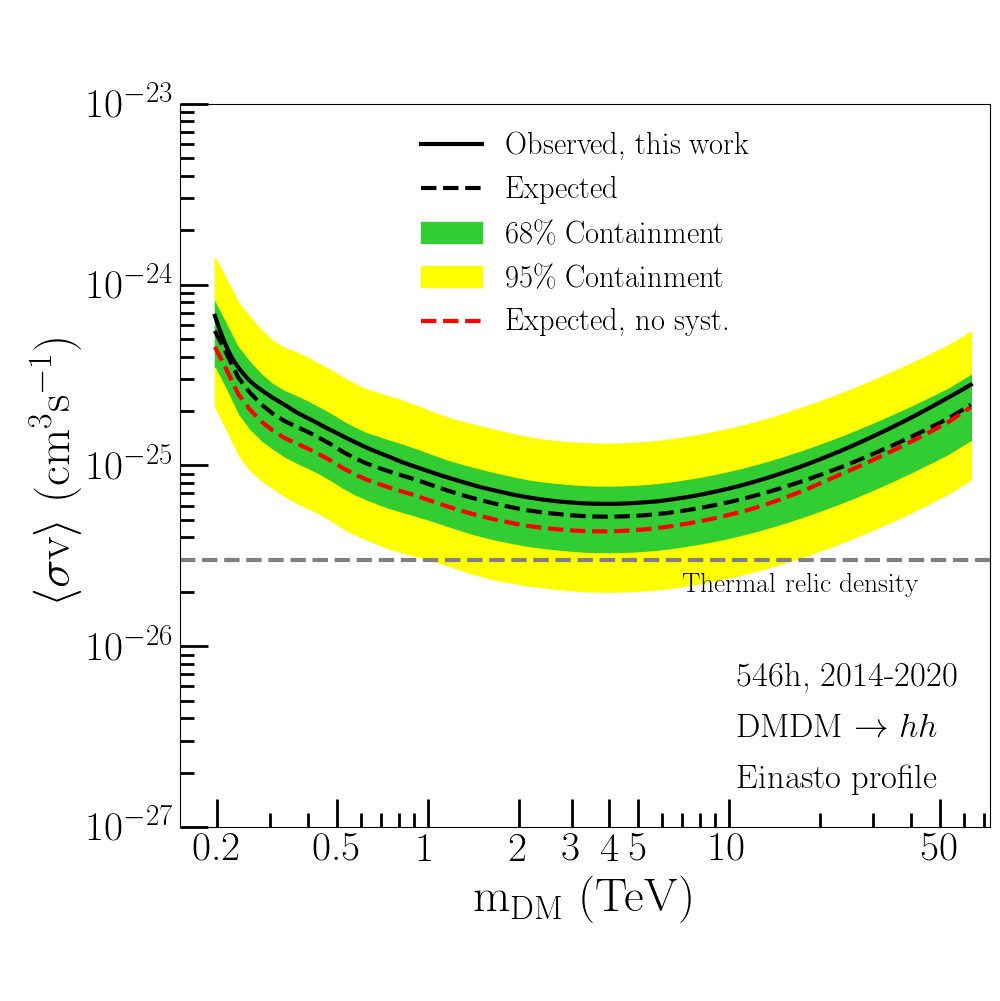}
\vspace{-0.5cm}
\includegraphics[width=0.45\textwidth]{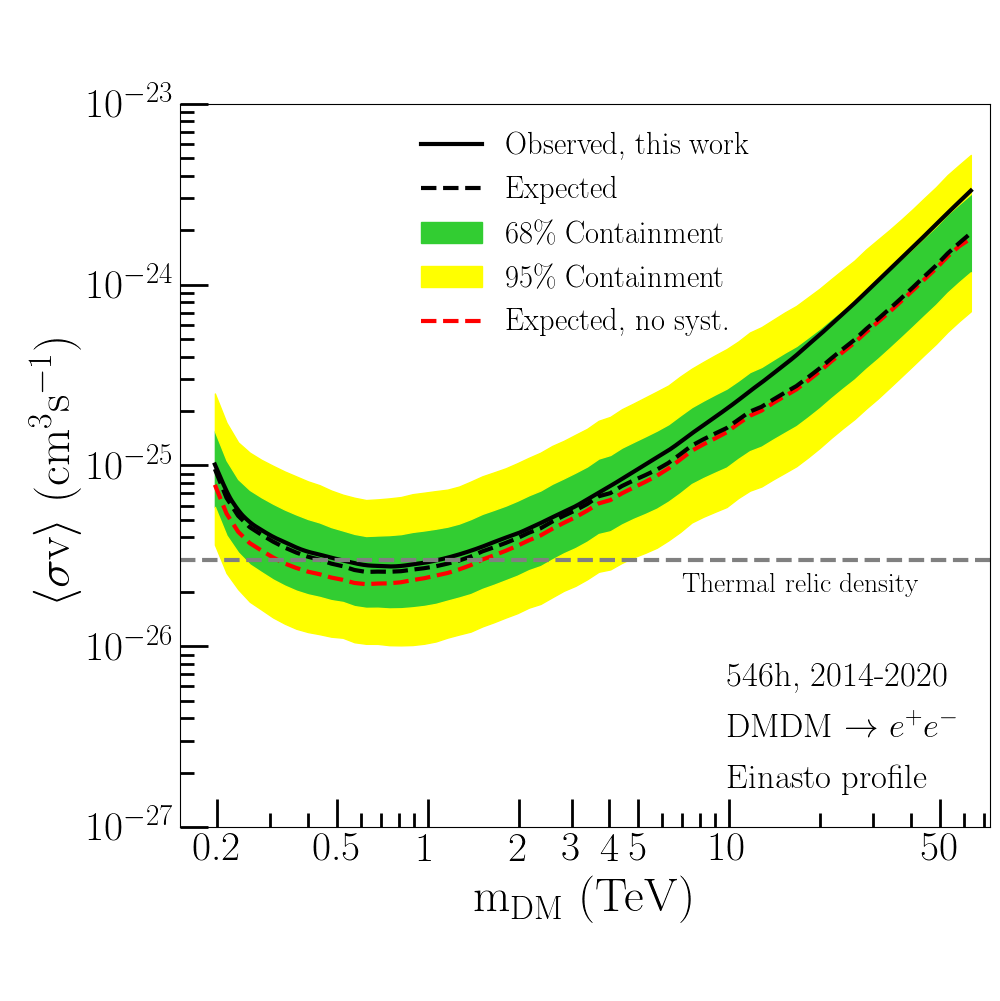}
\includegraphics[width=0.45\textwidth]{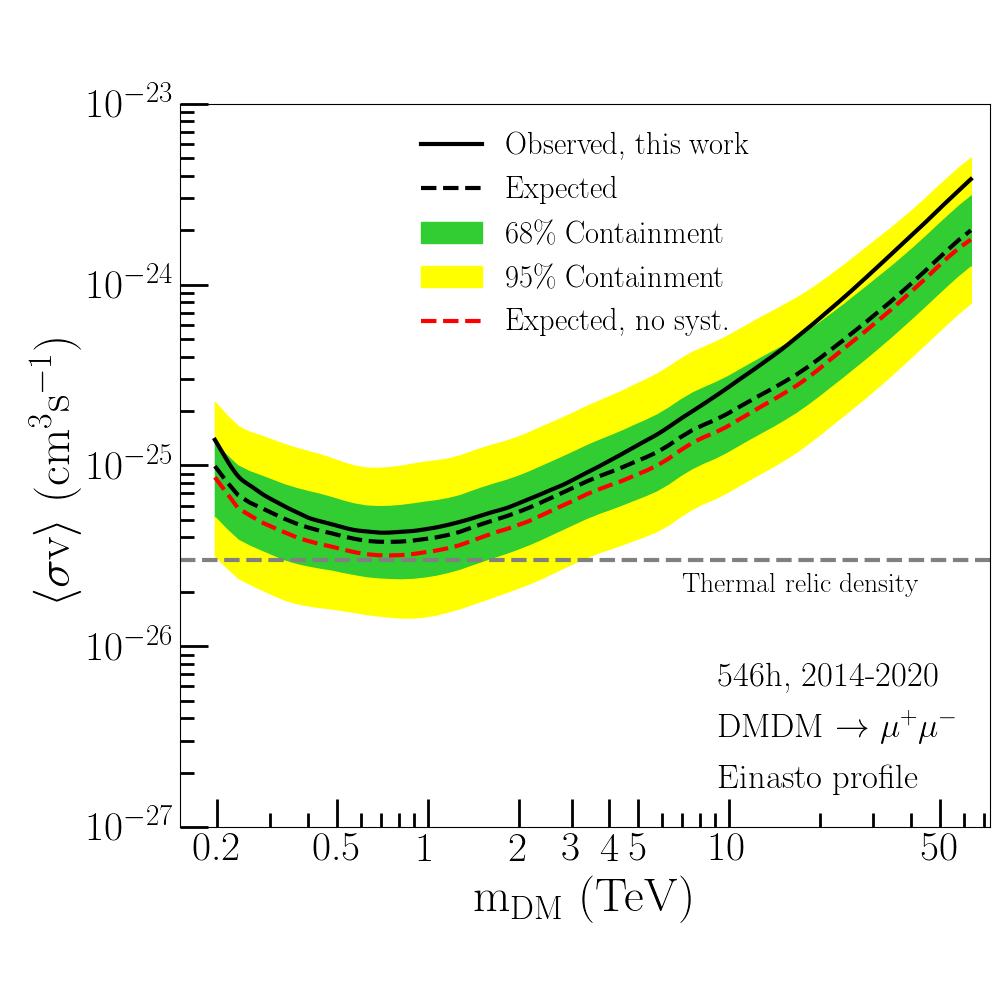}
\caption{Constraints on the velocity-weighted annihilation cross section $\langle \sigma v \rangle$ 
for the $b\bar{b}$, $t\bar{t}$, $ZZ$, $hh$, $e^+e^-$ and $\mu^+\mu^-$ channels, respectively, 
derived from H.E.S.S. five-telescope observations  taken from 2014 to 2020. The constraints are given as 95\% C. L. upper limits including the systematic uncertainty, as a function of the DM mass m$_{\rm DM}$.
The observed limit is shown as black solid line. 
The mean expected limit (black dashed line) together with the 68\% (green band) and 95\% (yellow band) C. L. containment bands are shown. 
The mean expected upper limit without systematic  uncertainty is also plotted 
(red dashed line). The horizontal grey long-dashed line is set to the value of the natural scale expected for the thermally-produced WIMPs.}
\label{fig:results_channels}
\end{figure*}

WIMPs can self-annihilate into pairs of Standard Model particles allowed by kinematics, providing gamma rays in the final states from hadronization, decay and radiation of the particles produced in the annihilation process.
We perform the analysis in the 
$b\bar{b}$, $t\bar{t}$, $W^+W^-$, $ZZ$, $hh$, $e^+e^-$, $\mu^+\mu^-$, and $\tau^+\tau^-$ annihilation channels, assuming a 100\% branching ratio in each case. The constraints are shown in Fig.~\ref{fig:results_channels} for the $b\bar{b}$, $t\bar{t}$, $ZZ$, $hh$, $e^+e^-$ and $\mu^+\mu^-$ channels, respectively.

% \end{document}

\end{document}